\documentclass[sort&compress,final,5p,twocolumn,times,fixfloat]{elsarticle}

\journal{Physics Letters B}

\usepackage{color}
\usepackage[dvipsnames]{xcolor}
\usepackage{bm}
\usepackage{amsmath}
\usepackage{amssymb}
\usepackage{booktabs}
\usepackage{float}
\usepackage[caption=false]{subfig} 
\usepackage{tabularx}
\usepackage{scalerel}
\usepackage{tikz}
\usetikzlibrary{svg.path}

\definecolor{orcidlogocol}{HTML}{A6CE39}

\usepackage[hidelinks]{hyperref}
\usepackage[sort&compress, capitalise]{cleveref}
 \hypersetup{
   pdftitle={},%
   pdfauthor={},%
   pdfsubject={},%
   pdfkeywords={},%
   pdfstartview={},%
   bookmarksopen=true, breaklinks=true, debug=true, %
   colorlinks=true, linkcolor=red, citecolor=blue, urlcolor=blue
 }

\usepackage{comment}

\newcommand{\eg}{{\it e.g.~}}

\newcommand{\itk}{\mathit{k}}
\newcommand{\itl}{\mathit{l}}

\newcommand{\wk}{\mathit{w}_\itk}
\newcommand{\wl}{\mathit{w}_\itl}

\newcommand{\wkl}{\mathit{w}_{\itk\itl}}
\newcommand{\bfa}{\boldsymbol{a}}
\newcommand{\bfb}{\boldsymbol{b}}
\newcommand{\BF}[1]{{\boldsymbol{#1}}}

\newcommand{\bfk}{\boldsymbol{k}}

\usepackage[normalem]{ulem}

\DeclareRobustCommand{\sec}[1]{Section~\ref{sec:#1}}

\newcommand*\oline[1]{%
   \vbox{%
     \hrule height 0.5pt
     \kern0.4ex
     \hbox{%
       \kern-0.15em
       \ifmmode#1\else\ensuremath{#1}\fi
       \kern-0.15em
     }
   }
}

\usepackage{ulem}

\begin{document}

\title{Simultaneous reweighting of Transverse Momentum Dependent distributions}

\date{\today}

\author[add1,add2]{M.~Boglione}
\ead{elena.boglione@to.infn.it}

\author[add3,add4]{U.~D'Alesio}
\ead{umberto.dalesio@ca.infn.it}

\author[add1,add2]{C.~Flore\corref{cor1}}
\ead{carlo.flore@to.infn.it}
\cortext[cor1]{Corresponding author}

\author[add1,add2]{J.~O.~Gonzalez-Hernandez}
\ead{joseosvaldo.gonzalezhernandez@unito.it}

\author[add4]{F.~Murgia}
\ead{francesco.murgia@ca.infn.it}

\author[add5,add6]{A.~Prokudin}
\ead{prokudin@jlab.org}

\address[add1]{Dipartimento di Fisica Teorica, Universit\`a di Torino, Via P.~Giuria 1, Torino, I-10125, Italy}
\address[add2]{INFN, Sezione di Torino, Via P.~Giuria 1, Torino, I-10125, Italy}
\address[add3]{Dipartimento di Fisica, Universit\`a di Cagliari, Cittadella Universitaria, I-09042 Monserrato (CA), Italy}
\address[add4]{INFN, Sezione di Cagliari, Cittadella Universitaria, I-09042 Monserrato (CA), Italy}
\address[add5]{Division of Science, Penn State University Berks, Reading, Pennsylvania 19610, USA}
\address[add6]{Theory Center, Jefferson Lab, 12000 Jefferson Avenue, Newport News, Virginia 23606, USA}

\begin{abstract}
The Bayesian reweighting procedure is extended to the case of multiple  independent extractions of transverse momentum dependent parton distributions (TMDs). By exploiting the data on transverse single spin asymmetries, $A_N$, for inclusive pion production in polarized proton-proton collisions measured at RHIC, we perform a simultaneous reweighting of the quark Sivers, transversity and Collins TMD functions extracted from semi-inclusive deep inelastic scattering (SIDIS) 
and $e^+ e^-$ annihilation into hadron pairs. The impact of the implementation of the Soffer bound, as well as the differences between older and newer $A_N$ data, are investigated. The agreement with $A_N$ data at large-$x_F$ values, a kinematical region complementary to those explored in SIDIS measurements, is enhanced, improving the knowledge of the polarized quark TMDs in the large-$x$ region.
\end{abstract}


\maketitle

\section{\label{sec:introduction} Introduction}

The idea of incorporating intrinsic transverse motion into the parton distribution functions dates back to the papers by Feynman, Field, and Fox who proposed to use it for the description of the transverse momentum dependent Drell-Yan cross-sections~\cite{Feynman:1977yr,Feynman:1978dt}. These functions were later named Transverse Momentum Dependent distribution and fragmentation functions (TMD-PDFs and TMD-FFs), collectively referred to as TMDs, and the TMD formalism was developed for (polarized) Semi-Inclusive Deep Inelastic Scattering (SIDIS), Drell-Yan, and $e^+e^-$ annihilation into hadron pairs~\cite{Kotzinian:1994dv,Tangerman:1994eh,Tangerman:1995hw,Boer:1997nt,Boer:1997mf,Bacchetta:2006tn,Arnold:2008kf,Pitonyak:2013dsu}. QCD factorization theorems were developed for TMDs~\cite{Collins:1981uk,Collins:1984kg, Ji:2004wu, Collins:2011zzd} and they were probed experimentally in various processes~\cite{HERMES:2009lmz,COMPASS:2008isr,Belle:2008fdv,COMPASS:2012dmt,BaBar:2013jdt,JeffersonLabHallA:2011ayy}. TMD physics is one of the pillars of the experimental programs of JLab 12~\cite{Dudek:2012vr} and the future Electron-Ion Collider (EIC)~\cite{Boer:2011fh,Accardi:2012qut}, as well as of RHIC~\cite{Aschenauer:2015eha} at BNL, COMPASS/AMBER~\cite{COMPASS:2010shj,Bradamante:2018ick,Adams:2018pwt} at CERN, BABAR~\cite{BaBar:2013jdt} at SLAC, Belle II~\cite{Belle:2008fdv} at KEK and BESIII~\cite{BESIII:2015fyw} in Beijing, of the fixed-target program at the LHC~\cite{Aidala:2019pit} and at Tevatron with the SpinQuest~\cite{SeaQuest:2019hsx} Drell-Yan program.

Historically, TMDs played a crucial role in explaining spin asymmetries~\cite{Anselmino:1994tv,Anselmino:1999pw,Boglione:1999dq,Anselmino:2004nk,DAlesio:2004eso,Anselmino:2005sh,DAlesio:2007bjf} and, in particular, the large value of the so-called left-right ($A_N$) or single-spin asymmetry (SSA) observed in proton-proton collisions~\cite{Bunce:1976yb,Klem:1976ui,E581:1991eys,Krueger:1998hz,Allgower:2002qi,STAR:2003lxu,PHENIX:2005jxc,Lee:2007zzh,STAR:2008ixi,BRAHMS:2008doi,STAR:2012hth,STAR:2012ljf,AnDY:2013att,PHENIX:2013wle,PHENIX:2014qwb,PHENIX:2020sqa,STAR:2020nnl}. Later on, it was argued that the so-called twist-3 formalism~\cite{Qiu:1991pp,Qiu:1998ia} is appropriate for the description of $A_N$, and it was shown that TMD factorization is, at least formally, violated in hadron production in proton-proton collisions~\cite{Rogers:2010dm}. Nevertheless, TMD and twist-3 formalisms are intimately connected~\cite{Ji:2006ub}, and TMD and twist-3 functions can be related either via specific integral expressions~\cite{Boer:2003cm,Gamberg:2017jha,Qiu:2020oqr,Ebert:2022cku,Gonzalez-Hernandez:2023iso,delRio:2024vvq} or through an operator product expansion~\cite{Collins:2011zzd,Vladimirov:2021hdn,Rein:2022odl}. Recently it was demonstrated~\cite{Cammarota:2020qcw,Gamberg:2022kdb} that spin asymmetries can be successfully fitted using TMD and twist-3 formalisms.

By extending our previous study~\cite{Boglione:2021aha}, in this paper we will attempt, for the first time, a simultaneous analysis of the available experimental data for spin asymmetries in SIDIS, $e^+e^-$ scattering, and proton-proton collisions, assuming factorized expressions in terms of TMDs for all those processes. We will exploit two models for the TMD description of $A_N$: the usual Generalized Parton Model (GPM)~\cite{DAlesio:2004eso,Anselmino:2005sh,DAlesio:2007bjf} which assumes that all TMDs are universal, and the Color Gauge Invariant  GPM (CGI-GPM)~\cite{Gamberg:2010tj,DAlesio:2011kkm, DAlesio:2017rzj,DAlesio:2018rnv} that takes into account the process dependence of TMDs due to the direction of gauge links in their corresponding operator definitions. The study will be performed by extending the Bayesian reweighting technique~\cite{Giele:1998gw,Ball:2010gb,Sato:2013ika,Sato:2016wqj,Lin:2017stx,DAlesio:2020vtw} to simultaneously reweight the results of new and updated global extractions of the transversity and Sivers distribution functions~\cite{Sivers:1989cc,Sivers:1990fh} and of the Collins fragmentation functions (FFs)~\cite{Collins:1992kk}, using presently available data on $A_N$ in proton-proton collisions.

The paper is organized as follows: in \sec{formalism} we recall the TMD formalism, within the GPM and CGI-GPM, while in \sec{simultaneous-reweighting} we summarize the basics of the reweighting procedure. A suitable method to treat Monte Carlo sets is discussed in \sec{compression}. The new independent fits to SIDIS and $e^+e^-$ data are presented in \sec{priors}, while the results of our analysis are discussed in \sec{results}. Conclusions and final remarks are gathered in \sec{conclusions}.

\section{\label{sec:formalism} Formalism}

In this Section we summarize the formalism which will guide us throughout our phenomenological analysis.

Starting with the SIDIS processes, $\ell p^\uparrow  \to \ell^\prime h X$, the two azimuthal asymmetries we are interested in are related to the Sivers and the Collins effects, properly defined within a TMD factorization theorem. For the Sivers asymmetry we have~\cite{Bacchetta:2004jz}
\begin{equation}
\label{eq:siv-asy}
A_{UT}^{\sin(\phi_h-\phi_S)} = \frac{F_{UT}^{\sin(\phi_h-\phi_S)}}{F_{UU}} \,,
\end{equation}
where $F_{UU}\sim f_1^q \otimes D_1^q$ is the TMD unpolarized structure function, and $F_{UT}^{\sin(\phi_h-\phi_S)}\sim f_{1T}^{\perp q} \otimes D_1^q$~\cite{Kotzinian:1994dv,Mulders:1995dh,Bacchetta:2006tn} is the azimuthal modulation originating from the correlation between the nucleon spin and the intrinsic transverse momentum of the unpolarized quark. This effect is encoded in the Sivers function. 

For the Collins asymmetry, which involves both transversity and Collins functions, one has
\begin{equation}
    A_{UT}^{\sin( \phi_h+\phi_S)} = \frac{2 (1-y)}{1+(1-y)^2}\frac{F_{UT}^{\sin(\phi_h+\phi_S)}}{F_{UU}} \; ,
\end{equation}
where $y$ is the fractional energy loss of the incident lepton, and $F_{UT}^{\sin(\phi_h+\phi_S)}\sim h_1^q \otimes H_1^{\perp q}$~\cite{Kotzinian:1994dv,Mulders:1995dh,Bacchetta:2006tn} is the polarized structure function of the SIDIS cross section, given as a convolution of the TMD transversity distribution, $h_1^q$, and the Collins FF, $H_1^{\perp q}$.
To access this TMD fragmentation function, information from  another complementary process, namely $e^+e^-\to h_1 h_2 X$, is necessary. Here the transverse momentum imbalance of the two hadron, produced in opposite hemispheres, is measured. In this configuration, still within a TMD factorization scheme, a convolution of two Collins FFs appears via a $\cos (2 \phi_0)$ modulation~\cite{Boer:1997mf}:
\begin{equation}
A_0^{UL(C)}\sim H_1^{\perp {\bar q}} \otimes H_1^{\perp q}\,.
\end{equation}
Experimental measurements of this process were conducted at approximately  $\sqrt{s} \simeq 10.6$ GeV by the Belle~\cite{Belle:2008fdv} and BABAR~\cite{BaBar:2013jdt} collaborations, as well as by the BESIII~\cite{BESIII:2015fyw} collaboration, at a lower energy of $\sqrt{s} \simeq 3.65$ GeV.

For inclusive hadron production in $pp$ collisions, the SSA is defined as
\begin{equation}\label{eq:AN}
A_N = \frac{d\sigma^\uparrow-d\sigma^\downarrow}{d\sigma^\uparrow+d\sigma^\downarrow} =\, \frac{d\Delta\sigma}{ 2 d\sigma}\,,
\end{equation}
where $d\sigma^{\uparrow(\downarrow)} \equiv E_h \, d\sigma^{\uparrow(\downarrow)}/d^3\bm{P}_h$  stands for the single-polarized cross section, in which one of the initial-state protons is transversely polarized ($\uparrow$($\downarrow$)) with respect to the production plane.
Here, we adopt the GPM, a phenomenological model where a factorized formulation in terms of TMDs is assumed as the starting point, and in which one includes spin and transverse momentum correlation effects. For completeness, we will also consider an extension of this approach, the CGI-GPM, where initial- and final-state interactions are properly included in a one-gluon-exchange approximation. Note that this model allows for the well-known process dependence of the Sivers function expected when 
comparing SIDIS and Drell-Yan processes~\cite{Collins:2002kn}.

As discussed in Refs.~\cite{Anselmino:2005sh,DAlesio:2007bjf}, in the region of moderate and forward rapidity in $p^\uparrow p \to h X$ processes only two effects survive the integration over the intrinsic transverse momenta and their relative azimuthal phases: the Sivers and Collins effects. In the first case, formally, also a contribution from gluons could appear. Nonetheless, in the same kinematical region, the gluon Sivers effect can safely be ignored, as shown in Refs.~\cite{DAlesio:2015fwo, DAlesio:2018rnv}.

It is important to stress that in inclusive processes these two TMD effects cannot be separated. Therefore the numerator of $A_N$ will be
\begin{equation}
d\Delta \sigma \simeq d\Delta \sigma_{\rm Siv} + d\Delta \sigma_{\rm Col}\,.
\end{equation}

Starting with the Sivers effect, within the CGI-GPM, the numerator of the asymmetry can be schematically written as~\cite{Gamberg:2010tj}
\begin{eqnarray}
\label{eq:sivgen}
d\Delta\sigma_{\rm Siv}^{\rm CGI-GPM} 
& \propto & \sum_{a,b,c,d} f^{\perp a}_{1 T}(x_a, k_{\perp a})\cos\varphi_a \otimes 
 f_{b/p}(x_b, k_{\perp b}) \nonumber \\
& & \mbox{}\otimes H^{\rm Inc}_{ab \to cd}\otimes D_{h/c}(z, {k}_{\perp h})\,,
\end{eqnarray}
where $f_{b/p}(x_b, k_{\perp b})$ is the TMD distribution for an unpolarized parton $b$ inside the unpolarized proton. Moreover, $H^{\rm Inc}_{ab \to cd}$ are the perturbatively calculable hard scattering functions. In particular, the $H^{\rm Inc}_{ab \to cd}$ functions where $a$ is a quark or an antiquark can be found in Ref.~\cite{Gamberg:2010tj}. The GPM result can be obtained from Eq.~(\ref{eq:sivgen}) by simply replacing $H^{\rm Inc}_{ab \to cd}$ with the standard tree-level unpolarized partonic cross sections, $H^U_{ab \to cd}$.
Finally, the unpolarized cross section, $d\sigma$, appearing in the denominator of~\cref{eq:AN}, can be obtained by replacing the Sivers function and its phase in the GPM expression with the corresponding unpolarized TMD-PDF for parton $a$.

Focusing now on the Collins contribution, we recall that all FFs (T-even as well as T-odd ones) are process independent, and are not modified by the direction of the gauge links~\cite{Collins:2004nx,Yuan:2009dw}. Thus, the Collins contribution to $A_N$ is assumed to be the same in the GPM and in the CGI-GPM, and reads
\begin{eqnarray}
\label{eq:colgen}
 d\Delta\sigma_{\rm Col}
&\propto &\sum_{a,b,c,d} h_{1 a}(x_a, k_{\perp a})\otimes 
 f_{b/p}(x_b, k_{\perp b}) \nonumber \\
&& \mbox{} \otimes d\Delta\sigma^{a^\uparrow b \to c^\uparrow d}\otimes H_{1}^{\perp c}(z, k_{\perp h}),
\end{eqnarray}
where $d\Delta\sigma^{a^\uparrow b \to c^\uparrow d} \equiv d\sigma^{a^\uparrow b \to c^\uparrow d} -d\sigma^{a^\uparrow b \to c^\downarrow d}$ is the transverse spin transfer at the partonic level. 

\section{\label{sec:simultaneous-reweighting} Simultaneous reweighting}

We now illustrate the method we have developed for a simultaneous Bayesian reweighting of functions initially extracted from fits to independent datasets. For simplicity, we will focus on the case of {\it two} functions,
although this approach can be easily generalized to $n$ independent extractions. 

Let us consider two statistically independent functions, $f(\BF{a})$ and $g(\BF{b})$ depending, respectively, on $n_a$- and $n_b$-dimensional sets of parameters $\BF{a} = \left\{a_1, \dots, a_{n_a}\right\}$ and $\BF{b} = \{b_1, \dots, b_{n_b}\}$. The value of these parameters is determined by performing  two distinct fits to independent datasets $\bm{E}^a$ and $\bm{E}^b$. For each of these fits, a $\chi^2$, defined as~\footnote{If (\eg for the fit $\bfa$) only uncorrelated uncertainties $\sigma_i^a$ are given, the new $\chi^2$ reduces simply to $\chi^2_a[\bfa; \bm{E}^a]= \sum\limits_{i = 1}^{N_{\rm dat}^a} \dfrac{(T_i[\bfa] - E_i^a)^2}{({\sigma_i^a})^2}$.}{$^,$}\footnote{In what follows the indices ($i$,$j$) will be used for individual data points, while the indices ($k$,$l$) will refer to MC sets.}:
\begin{equation}
\begin{aligned}
    & \chi^2_a \equiv \chi^2 [\bfa; {\bm E}^a] = \sum_{i,j = 1}^{N_{{\rm dat}}^a} (T_i[\bfa] - E_i^a)\, (C_{ij}^{a})^{-1}(T_j[\bfa] - E_j^a)\,, \\
    & \chi^2_b \equiv \chi^2 [\bfb; {\bm E}^b] = \sum_{i,j = 1}^{N_{{\rm dat}}^b} (T_i[\bfb] - E_i^b)\, (C_{ij}^{b})^{-1}(T_j[\bfb] - E_j^b)\,,
\end{aligned}
\end{equation}
is minimized and the best fit $\bfa_0$ and $\bfb_0$, corresponding to the minima $\chi^2_{0,a}$ and $\chi^2_{0,b}$, are determined. In the equations above, $T_i[\bfa] \equiv T_i(f(\bfa))$ are the theoretical estimates corresponding to the experimental data points $E_i^a$, and $C_{ij}^a$ is the covariance matrix for the fit $\bfa$ (and similarly for the fit $\bfb$). The fit uncertainties can then be computed via a Hessian method or with a suitable Monte Carlo (MC) procedure. Using the latter method, the probability density functions $\pi(\bfa)$ and $\pi(\bfb)$ are reconstructed by generating $N_{\rm set}^a$ sets $\bfa_k$ and $N_{\rm set}^b$ sets $\bfb_l$ respectively. Notice that these distributions are statistically independent from each other. Then, expectation values and variances for any quantity $\mathcal{O}$ depending on one of the parameter sets (\textit{e.g.}~$\bfa$) can be computed as
\begin{equation}\label{eq:E-V-unweighted}
\begin{aligned}
 &{\rm E}[{\mathcal O}] = \frac{1}{N_{\rm set}^a} \sum\limits_{\itk = 1}^{N_{\rm set}^a} {\mathcal O}(\bfa_{\itk})\,, \\
 &{\rm V}[\mathcal{O}] = \frac{1}{N_{\rm set}^a} \sum\limits_{\itk = 1}^{N_{\rm set}^a} \left(\mathcal{O}(\bfa_\itk) - {\rm{E}}[\mathcal{O}]\right) ^2\,.
\end{aligned}
\end{equation}

Let us now suppose that a new set of data ${\bm E}$ (with an associated covariance matrix $C$) is measured, and that these data can be described by a linear combination of $f(\bfa)$ and $g(\bfb)$ (\textit{e.g.}~$T_i[\bfa,\bfb]\equiv \alpha T_i[\bfa] + \beta T_i[\bfb]$, where $\alpha$ and $\beta$ are real constants). Then, we can compute the $\chi^2$ corresponding to these new data as
\begin{equation}
    \chi^2_{{\rm new}}[\bfa, \bfb; {\bm E}] = \sum_{i,j = 1}^{N_{\rm dat}} (T_i[\bfa, \bfb] - E_i)\, C_{ij}^{-1}(T_j[\bfa, \bfb] - E_j)\,.
\end{equation}

Since $f$ and $g$ come from statistically independent fits, the uncertainty bands for the theoretical predictions $T_i[\bfa, \bfb]$ have to be built by taking all possible ($N_{\rm set}^a\times N_{\rm set}^b$) combinations of the MC parameter sets $\bfa_k$ and $\bfb_l$. Thus, the $\chi^2$ on the new data will depend on the $k$-th and $l$-th MC sets:
\begin{equation}
\label{eq:chi2-matrix}
    \chi^2_{{\rm new}} \equiv \chi^2_{kl,{\rm new}} = \chi^2_{{\rm new}}[\bfa_k, \bfb_l; {\bm E}]
\end{equation}
leading to ($N_{\rm set}^a\times N_{\rm set}^b$) values of $\chi^2$.

By using Bayes theorem, we can then evaluate the impact of these new data on our prior distributions $\pi(\bfa)$ and $\pi(\bfb)$. Since these distributions are \textit{a priori} independent, we can build a factorized prior $\pi(\bfa, \bfb) = \pi(\bfa)\pi(\bfb)$ and apply Bayes theorem to compute the posterior densities:
\begin{equation}
   {\cal P}(\bfa, \bfb | \BF{E} )=\frac{{\cal L} (\BF{E} | \bfa, \bfb)\, \pi(\bfa, \bfb)}{\mathit{Z}}\;,
   \label{eq:posterior}
\end{equation}
where ${\cal L} (\BF{E}| \bfa, \bfb)$ is the likelihood, and $Z = {\cal P}({\BF{E}})$ is the evidence, that ensures a normalized posterior density.

Various choices for the likelihood have been discussed in the literature~\cite{Giele:1998gw,Ball:2010gb}. Here we adopt the likelihood definition as obtained by taking ${\cal L} (\BF{E}| \bfa, \bfb)\,d{\BF{E}}$ as the probability to find the new data confined in a differential volume $d\BF{E}$ around $\BF{E}$. Following Ref.~\cite{Paukkunen:2014zia}, we define the weights $\wkl$ as
\begin{equation}\label{eq:weights}
 \wkl (\chi^2_{\rm new}) = \frac{{\rm exp}\left\{-\frac12\frac{\chi^2_{kl, {\rm new}}}{\Delta\chi^2}\right\}}{\sum\limits_{k^\prime,l^\prime} {\rm exp}\left\{-\frac12\frac{\chi^2_{k^\prime l^\prime, {\rm new}}}{\Delta\chi^2}\right\} }\,,
\end{equation}
where $\Delta\chi^2$ is the tolerance at a given confidence level (CL) for $n_a+n_b$ parameters. Notice that the weights coincide with those defined in the original work by Giele and Keller~\cite{Giele:1998gw}, with rescaled exponent: $\chi^2_{kl} \to \chi^2_{kl} / \Delta\chi^2$. We will use a value of $\Delta\chi^2$ defined according to Wilks' theorem~\cite{Wilks:1938dza}. For a $(1-\alpha)$ CL, we have
\begin{equation}\label{eq:delta-chi2}
\Delta\chi^2=F^{-1}_{X^2_D}(1-\alpha)\,,        
\end{equation}
where $X_D^2$ is a chi-squared probability density for $D$ degrees of freedom (\textit{i.e.}~the number of free parameters) and $F_{X_D}^2$ its associated cumulative function.

The weights are at the core of the reweighting procedure. Using Eq.~\eqref{eq:weights}, one can obtain the expectation value and variance of the reweighted quantity $\mathcal{O}$ as
\begin{equation}\label{eq:E-V-reweighted}
\begin{aligned}
 \rm{E}[\mathcal{O}] & = \sum\limits_{k=1}^{N_{\rm set}^a}\sum\limits_{l=1}^{N_{\rm set}^b} \wkl\, \mathcal{O}(\bfa_\itk,\bfb_\itl)\,, \\
 \rm{V}[\mathcal{O}] & =\sum\limits_{k=1}^{N_{\rm set}^a}\sum\limits_{l=1}^{N_{\rm set}^b} \wkl \left(\mathcal{O}(\bfa_\itk,\bfb_\itl) - {\rm{E}}[\mathcal{O}]\right)^2\,.
\end{aligned}
\end{equation}
If this quantity depends only on $f(\bfa)$ (or $g(\bfb)$), one has to evaluate the corresponding weights $\wk$ (or $\wl$)
\begin{equation}
    \wk = \sum\limits_{l=1}^{N_{\rm set}^b} \wkl\,,\qquad \wl = \sum\limits_{k=1}^{N_{\rm set}^a} \wkl\,,
\end{equation}
and use again the weighted sums in Eq.~(\ref{eq:E-V-reweighted}) with the new, updated weights $\wk$ (or $\wl$) for the corresponding $\mathcal{O}(\bfa_\itk)$ (or $\mathcal{O}(\bfb_\itl)$). By doing so, one is able to evaluate the impact of the new data on the two independent prior distributions $\pi(\bfa)$ and $\pi(\bfb)$ and on any quantity depending on the parameter sets $\bfa$ and/or $\bfb$. As the weights defined in Eq.~(\ref{eq:weights}) are normalized to one, $\wk$ and $\wl$ are automatically normalized to one too.

For a generic extraction with $N_{\rm set}$ MC sets and weights $\wk$, the rescaled version is equivalent to the Hessian reweighting~\cite{Paukkunen:2014zia}, and allows us to retain a larger effective number of sets, $N_{\rm eff}$, defined as 
\begin{equation}\label{eq:N-eff}
    N_{\rm eff} = {\rm exp}\left\{\sum\limits_{k=1}^{N_{\rm set}}\wk \ln\left(\frac{1}{\wk}\right)\right\}\,.
\end{equation}
$N_{\rm eff}$ is related to the number of sets carrying a non-negligible weight, reflecting the method's efficiency. If $N_{\rm eff} \ll N_{\rm set}$, the method is considered no longer reliable, signaling that either the new data require a full refitting, or that they are inconsistent with the old ones~\cite{Ball:2010gb}.

In general, the introduction of new data may lead to correlations between fits that were originally statistically independent. For example, this scenario could arise with $A_N$, which incorporates contributions from both Sivers and Collins effects. Such correlations are encoded in the ($N_{\rm set}^a \times N_{\rm set}^b$) combinations, and are duly considered when evaluating a reweighted quantity.

\section{\label{sec:compression} Compressing the MC sets}

Following the procedure illustrated in Appendix A of Ref.~\cite{Anselmino:2008sga}, after the initial fitting stage we generate ({\it e.g.}~for the fit $\bfa$) $N^a_{\rm set}$ MC sets $\bfa_k$, each with a corresponding $\chi^2_{a,k} \in [\chi^2_{0,a}, \chi^2_{0,a} + \Delta\chi^2_a]$. Again, $\chi^2_{0,a}$ is the minimum found by the fit and $\Delta\chi^2_a$ is the tolerance that depends on the number of parameters and is given at a certain CL. These sets allow us to reliably reconstruct the parameter distribution $\pi(\bfa)$, provided a sufficiently large number of sets is generated. For instance, in Refs.~\cite{Boglione:2018dqd, DAlesio:2020vtw, Boglione:2021aha}, the number of sets needed was up to $O(10^6)$. In the case we consider here, involving two independent functions, this implies up to $O(10^{12})$ combinations in Eq.~(\ref{eq:chi2-matrix}). In order to reduce the computational cost, we will use a compression procedure, which we describe in what follows.

Starting from the full sample of parameters $\bfa_k$, we select a \textit{random} sample $\bfa_{k^\prime}^\prime = \{\bfa_1^\prime, \dots, \bfa^{\prime}_{N^{a^\prime}_{\rm set}}\}$ with $N^{a^\prime}_{\rm set} \ll N^a_{\rm set}$. If $\pi({\bfa_{k^\prime}^\prime}) \simeq \pi(\bfa_k)$, then one also expects that $\pi({\mathcal O}(\bfa_{k^\prime}^\prime)) \simeq \pi({\mathcal O}(\bfa_k))$. In other words, if the sample $\bfa_{k^\prime}^\prime$ renders a statistically equivalent distribution to that of $\bfa_k$, the corresponding distributions of any quantity will not differ significantly. Thus, this procedure would help in decreasing the number of combinations to be computed for the simultaneous reweighting. 

The question now is how to check that the sampled distribution is statistically equivalent to the full one.  As discussed in Section C.2 of Ref.~\cite{Bauer:2022mvl}, we adopt as an indicator the Welch's $t$-statistic, defined as
\begin{equation}\label{eq:welch-t}
t = \frac{\mu_{\bfa} - \mu_{\bfa^\prime}}
{\sqrt{\frac{\sigma^2_a}{N^a_{\rm set}} + \frac{\sigma^2_{a^\prime}}{N^{a^\prime}_{\rm set}}}}\,,
\end{equation}
which quantifies the difference of the arithmetical means of two samples (in our case, the full sample $\bfa_k$ and the compressed random sample $\bfa_{k^\prime}^\prime$) with unequal variances and sizes. One has to verify that $|t|$ is such that the corresponding $p$-values are $\gtrsim 0.1$. Provided this condition holds, one can conclude that the sampled distribution and the original distribution are statistically equivalent. Notice that, at variance with Ref.~\cite{Bauer:2022mvl}, we do not sample in the observable space (\textit{i.e.}~$A_N$ in our case), but rather in the parameter space. Moreover, since the Welch's $t$-statistic implicitly assumes underlying Gaussian distributions  we also verify the compatibility between the medians and \textit{asymmetric} uncertainty intervals of the samples $\bfa_{k^\prime}^\prime$ and $\bfa_k$. This allows us to correctly sample  asymmetric distributions. To check how this compression algorithm works, we detail below an explicit example.

\subsection{\label{sec:compression-example} An explicit example}

Let us consider the reweighting performed for the quark Sivers function using STAR $A_N$ jet data~\cite{Boglione:2021aha}. Here, we will re-perform the reweighting using the rescaled weights as defined in Eq.~(\ref{eq:weights}). 

In the original work, $N^a_{\rm set} = 2\cdot 10^5$ MC sets were generated and used to represent the uncertainty on the up- and down-quark Sivers functions. Here, we will select a random sample $\bfa_{k^\prime}^\prime$ of the parameter sets $\bfa_k$, with $N^{a^\prime}_{\rm set} \ll N^a_{\rm set}$, checking that their corresponding distribution $\pi(\bfa_{k^\prime}^\prime)$ is statistically equivalent to $\pi(\bfa_k)$ and re-perform the reweighting using only the reduced sample of sets. By applying the compression algorithm presented above, we select only 1\% of the initial sample, \textit{i.e.}~$N^{a^\prime}_{\rm set} = 2\cdot 10^3$ sets.

\begin{figure}[h!]
\centering
\vspace{-4mm}
\includegraphics[width=\columnwidth]{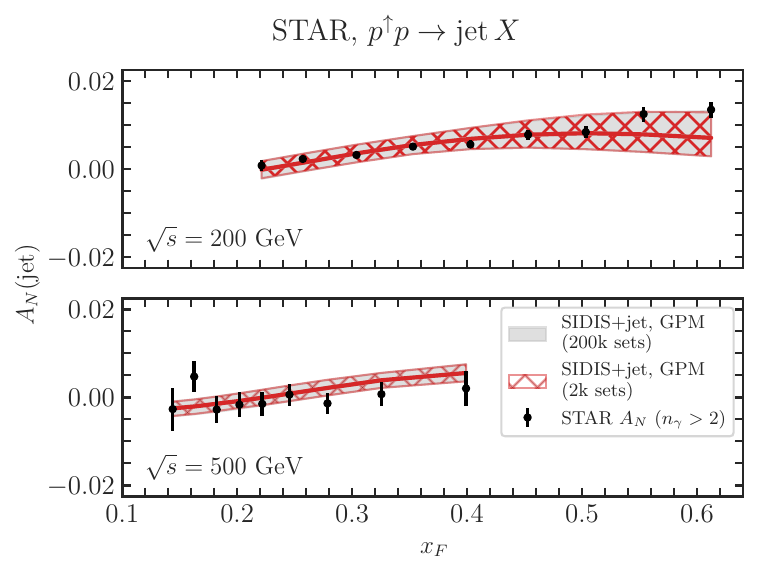}
\vspace{-7mm}
\includegraphics[width=\columnwidth]{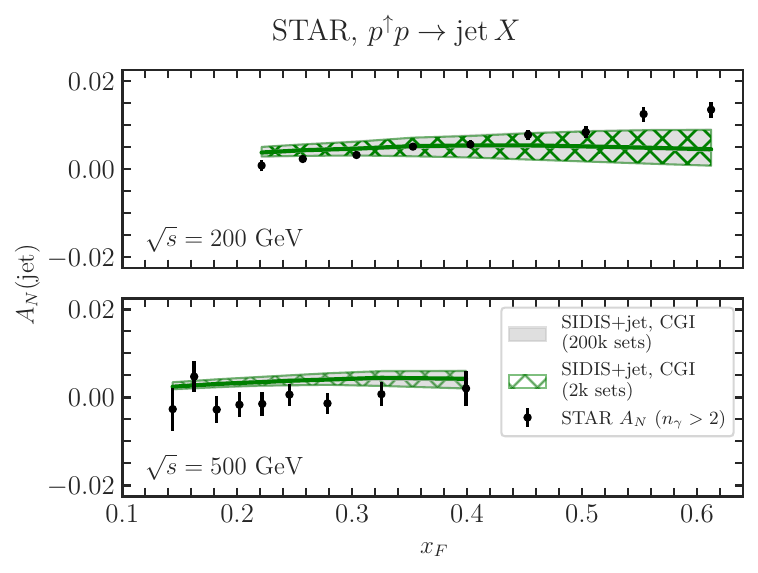}
\caption{Comparison between reweighted curves for the full (gray bands) and reduced (hatched bands) samples for the reweighting analysis of Ref.~\cite{Boglione:2021aha}.}
\label{fig:Siv-rew-200k-vs-2k}
\end{figure}
\begin{figure}[htbp]
\centering
\includegraphics[width=\columnwidth]{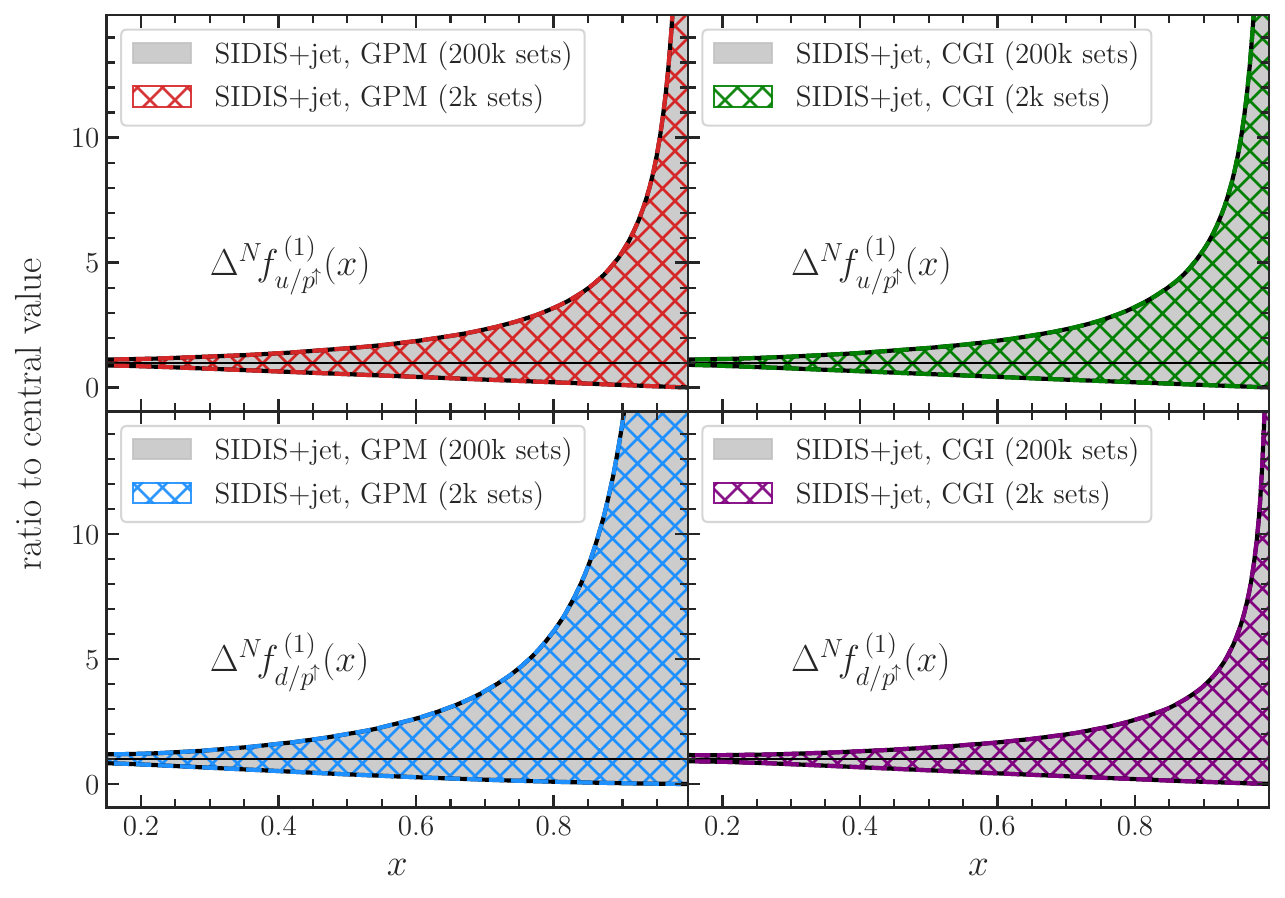}
\caption{Comparison between reweighted first moments of up- (upper panels) and down-quark (lower panels) Sivers functions, normalized to their central value, using full (gray bands) and reduced (hatched bands) samples for the reweighting analysis of Ref.~\cite{Boglione:2021aha}.}
\label{fig:Siv-ratios-200k-vs-2k}
\end{figure}
Fig.~\ref{fig:Siv-rew-200k-vs-2k} shows the comparison between reweighted curves for the GPM (upper panels) and the CGI-GPM (lower panels), together with STAR data. As in Ref.~\cite{Boglione:2021aha}, the central values are the median values, and the asymmetric uncertainty bands are at $2\sigma$ CL. The plot clearly shows that median values and uncertainties of the full sample (gray bands) are correctly and satisfactorily reproduced by the reduced sample of MC sets (hatched bands). We verified that the same happens for the unweighted curves, not shown here. For completeness, and to make this comparison more explicit, in Fig.~\ref{fig:Siv-ratios-200k-vs-2k} we show the reweighted uncertainties (normalized to their central value) for the Sivers first moment for up- and down-quark in the GPM (left panels) and CGI-GPM (right panels). Reweighted curves from the full sample are shown in gray with black dashed borders, while the ones for the reduced sample are shown in hatched colors. These results allow us to validate the compression procedure, that we will use in what follows for simultaneously reweighting the Sivers, transversity and Collins functions. 

\section{\label{sec:priors} Priors from SIDIS and $e^+e^-$ data}

In this Section we briefly describe the new fits to SIDIS and $e^+e^-$ data for the extraction of the Sivers, transversity and Collins functions. These will represent the priors for the simultaneous reweighting procedure. In all cases we employ updated SIDIS datasets, by including the most recent data from COMPASS~\cite{COMPASS:2017mvk}, HERMES~\cite{HERMES:2020ifk} and JLab~\cite{JeffersonLabHallA:2011ayy}.

The unpolarized TMD PDFs and FFs are parametrized using a factorized Gaussian ansatz:
\begin{equation}
\begin{aligned}
 & f_{a/p}(x, k_\perp^2) = f_{a/p}(x) \frac{e^{-k_\perp^2/\langle k_\perp^2\rangle}}{\pi \langle k_\perp^2\rangle} \\
 & D_{h/q}(z, p_\perp^2) = D_{h/q}(z)\, \frac{e^{-p^2_\perp / \langle p^2_\perp \rangle}}{\pi \langle p^2_\perp\rangle}
\end{aligned}
\end{equation}
with $\langle k_\perp^2\rangle= 0.57$ GeV$^2$ and $\langle p_\perp^2\rangle = 0.12$ GeV$^2$ as extracted from a fit to HERMES multiplicities~\cite{Anselmino:2013lza}. As collinear input, we adopt the MSHT20nlo proton PDFs~\cite{Bailey:2020ooq} and the DEHSS fragmentation functions for pions and kaons~\cite{deFlorian:2014xna,deFlorian:2017lwf}.  

For the up- and down-quark Sivers functions, we adopt the parametrization of Ref.~\cite{Boglione:2018dqd}, that consists in factorized $x$ and $k_\perp$ dependences (the latter being Gaussian-like and flavor independent):
\begin{equation}\label{eq:Sivers-parametrization}
    \Delta^N\!f_{q/p^\uparrow}(x,k_\perp) = \frac{4 M_p k_\perp}{\langle k_\perp^2\rangle_S} \Delta^N\!f_{q/p^\uparrow}^{(1)}(x) \frac{e^{-k_\perp^2/\langle k_\perp^2\rangle_S}}{\pi \langle k_\perp^2\rangle_S}\,,
\end{equation}
where $q = u, d$, $M_p$ is the proton mass, and where $\Delta^N\!f_{q/p^\uparrow}^{(1)}(x)$ is the Sivers first $k_\perp$-moment~\cite{Bacchetta:2004jz}:
\begin{equation}\label{eq:f1Tp-first-mom}
\begin{aligned}
    \Delta^N\!f_{q/p^\uparrow}^{(1)}(x)  & = \int d^2 \bfk_\perp \frac{k_\perp}{4 M_p} \Delta^N\!f_{q/p^\uparrow}(x,k_\perp) \equiv - f_{1T}^{\perp (1) q}(x) \\
    &= N_q\,(1-x)^{\beta_q}\,.
\end{aligned}    
\end{equation}
This model depends on five parameters: $N_u$, $N_d$, $\beta_u$, $\beta_d$, and $\langle k_\perp^2\rangle_S$. 

Following Section 3.1 of Ref.~\cite{Boglione:2018dqd}, in the computation of $F_{UU}$ (see Eq.~(\ref{eq:siv-asy})), we consistently employ the same collinear PDFs and FFs as for the unpolarized TMDs, and the corresponding Gaussian widths extracted from HERMES and COMPASS multiplicities~\cite{Anselmino:2013lza}. 

For the transversity and Collins functions we make use of the parametrization of Refs.~\cite{Anselmino:2015sxa, DAlesio:2020vtw}. The transversity function is parametrized as
\begin{equation}\label{eq:h1(x)-SB-Q0}
 h_1^q(x, k^2_\perp) = h_1^q (x) \frac{e^{-k^2_\perp / \langle k^2_\perp \rangle}}{\pi \langle k^2_\perp\rangle},
\end{equation}
where the Gaussian width is assumed to be the same as for the unpolarized TMD-PDFs. As in Refs.~\cite{Anselmino:2007fs, Anselmino:2013vqa, Anselmino:2015sxa, DAlesio:2020vtw}, the $x$-dependent part of the TMD transversity is parametrized at the initial scale $Q_0^2$ in terms of the Soffer bound~\cite{Soffer:1994ww}:
\begin{equation}\label{eq:h1(x)-SB}
\begin{aligned}
 h_1^q(x, Q_0^2) & = \mathcal{N}^T_q(x) \frac{1}{2} \left[f_{q/p}(x, Q_0^2) + g_{1L}^q(x, Q_0^2)\right]\\ & \equiv \mathcal{N}^T_q(x)\, {\rm SB}(x,Q_0^2),
\end{aligned}
\end{equation}
where
\begin{equation}
 {\cal N}^{T}_q(x)=N^{T}_q x^{\alpha}(1-x)^\beta\,
\frac{(\alpha+\beta)^{\alpha+\beta}}{\alpha^\alpha \beta^\beta},
\quad (q = u_v,\,d_v)
\end{equation}
with the same $\alpha$ and $\beta$ parameters for the valence $u_v$ and $d_v$ transversity functions, for a total of four parameters for $h_1^q$. We emphasize that we do not enforce the automatic fulfillment of the Soffer bound ($\lvert N_q^T\rvert \leq 1$), but we apply such a constraint \textit{a posteriori} on the generated MC sets. As shown in Ref.~\cite{DAlesio:2020vtw}, this choice allows us to avoid a bias in the fitting procedure and to properly estimate the uncertainty on the transversity functions.

The Collins function is parametrized as in Refs.~\cite{Anselmino:2007fs, Anselmino:2013vqa, Anselmino:2015sxa, DAlesio:2020vtw}:
\begin{equation}
\label{eq:Collins}
H_1^{\perp q}(z, p_\perp^2)  = {\cal N}^C_q (z) \frac{z m_h}{ M_C}\,\sqrt{2e}\,e^{-p_\perp^2/M_C^2}\,D_{h/q}(z, p_\perp^2)\,,
\end{equation}
where $q = \rm{fav}, \rm{unf}$ (favored/unfavored), $m_h$ is the produced hadron mass, and where $M_C$ is a free parameter with mass dimension. $D_{h/q}(z, p_\perp^2)$ is again the unpolarized TMD fragmentation function, while the ${\cal N}^C_q(z)$ factors are given by
\begin{equation}
\label{eq:Collins-NC}
 {\cal N}^C_{\rm{fav}}(z) = N^C_{\rm{fav}}\, z^\gamma, 
 \quad{\cal N}^C_{\rm{unf}}(z) = N^C_{\rm{unf}}\,,
\end{equation}
for a total of eight free parameters for the $h_1^q$ and $H_1^{\perp q}$ extraction.

To build the Soffer bound, we adopt the DSSV set~\cite{deFlorian:2009vb} for the collinear helicity distributions, $g_{1L}(x)$. By using an appropriately modified version~\cite{Courtoy:2012ry,Prokudin:hoppet} of the {\tt HOPPET} code~\cite{Salam:2008qg}, a transversity DGLAP kernel is employed to evolve $h_1(x)$ up to higher values of $Q^2$. We set $Q_0^2 = 0.81\,\text{GeV}^2$ as the input scale in Eq.~(\ref{eq:h1(x)-SB-Q0}), with $\alpha_S(M_Z) \simeq 0.118$. 

For the collinear part of the Collins function, we also adopt a DGLAP evolution. In principle scale evolution should be taken into account in a more rigorous way. In this case, in particular, the appropriate formalism would be that of TMD factorization leading to TMD evolution equations. There is, however, a general consensus based on experimental evidences that scale evolution effects appear to be mild when it comes to azimuthal or single-spin asymmetries. In fact, asymmetries are defined as  \emph{ratios} of cross sections, where evolution and higher order effects tend to cancel out~\cite{Kang:2017btw}. Although our parametrization does not incorporate the complete features of TMD evolution, phenomenological results based on DGLAP evolution are compatible with full TMD evolution at higher logarithmic accuracy~\cite{Kang:2017btw,DAlesio:2017bvu} (see also Fig.~14 of Ref.~\cite{STAR:2017akg}) in the kinematic region we are interested in.

Note that the updated extractions turn out to be compatible with those of Refs.~\cite{Anselmino:2013lza,Boglione:2018dqd,DAlesio:2020vtw}, although the new HERMES data induce slightly larger TMD distributions, as already observed in Ref.~\cite{Gamberg:2022kdb}. For the two independent extractions of TMDs from the Sivers and Collins asymmetries we generate $\mathit{O}(10^5)$ sets using a Markov chain MC that employs a Metropolis-Hastings algorithm with an auto-regressive generating density~\cite{10.2307/2684568}), and apply the compression algorithm discussed in \sec{compression} to select $2\cdot 10^3$ MC sets for each extraction. This amounts to $4\cdot 10^6$ combinations to be computed for each of the $A_N$ bins for the simultaneous reweighting.

As mentioned above, these updated analyses will represent the priors of the reweighting procedure, which will be described in the following Section.

\section{\label{sec:results} Results}

\subsection{Simultaneous reweighting with $A_N$ data for inclusive pion production}

We start by illustrating the comparison between our predictions for the $A_N$ asymmetry in inclusive production of charged and neutral pions in the GPM and CGI-GPM with the experimental data used for the simultaneous reweighting. We consider as new evidence the preliminary data for $A_N$ measured by BRAHMS for $\pi^\pm$ production at $\sqrt{s} = 200$ GeV~\cite{Lee:2007zzh}, the data from STAR for $\pi^0$ production at $\sqrt{s} = 200$ GeV~\cite{STAR:2003lxu,STAR:2008ixi,STAR:2012ljf} and the latest STAR data for non-isolated $\pi^0$ production from Ref.~\cite{STAR:2020nnl} at $\sqrt{s} = 200$ GeV and $\sqrt{s} = 500$ GeV. 

In our computation of $A_N$, the transverse momentum of the final state pion, $P_T$, is the hard scale of the process. For the (CGI-)GPM to be applicable, we then select only data points with $P_T > 1$ GeV.

In what follows we adopt the median as central value, and the uncertainties are estimated by determining $2\sigma$-confidence regions. Since we have a total of 13 free parameters (5 for $f_{1T}^\perp$, 4 for $h_1$ and 4 for $H_1^\perp$), according to Eq.~\eqref{eq:delta-chi2}, we get $\Delta\chi^2=22.69$ entering Eq.~\eqref{eq:weights}.

Hereafter, we present the unweighted predictions, based on the information from SIDIS and $e^+e^-$ asymmetries only, in gray. The reweighted curves in the GPM and the CGI-GPM are shown respectively with red and green bands. Data points corresponding to $P_T <1.5$ GeV are depicted in gray, to highlight the kinematic regions where the perturbative approach may be questioned, especially as far as scale uncertainties are concerned (see {\it e.g.}~Ref.~\cite{ColpaniSerri:2021bla} for a recent discussion on this issue).

Let us start from the results for charged pion production at BRAHMS. Before entering into our main discussion, few comments are in order. While it is well known that $\pi^0$ data are mostly sensitive to the relative contribution of up- and down-quark TMD distributions (Sivers or transversity, depending on the effect considered), $\pi^\pm$ data allow for a more direct flavor separation, giving a larger discriminating power to any phenomenological study. Therefore, in view of their relevance, we have  included the charged pion datasets in our analysis, although yet unpublished and covering a limited kinematical range. Charged pion $A_N$ measurements at future facilities, like the EIC~\cite{Boer:2011fh,Accardi:2012qut}, the JLab 22 program~\cite{Accardi:2023chb}, AMBER~\cite{Adams:2018pwt} and the proposed fixed-target program at the LHC~\cite{Aidala:2019pit}, will indeed help in improving future TMD analyses.

\begin{figure}[h!]
\centering
\includegraphics[width=\columnwidth]{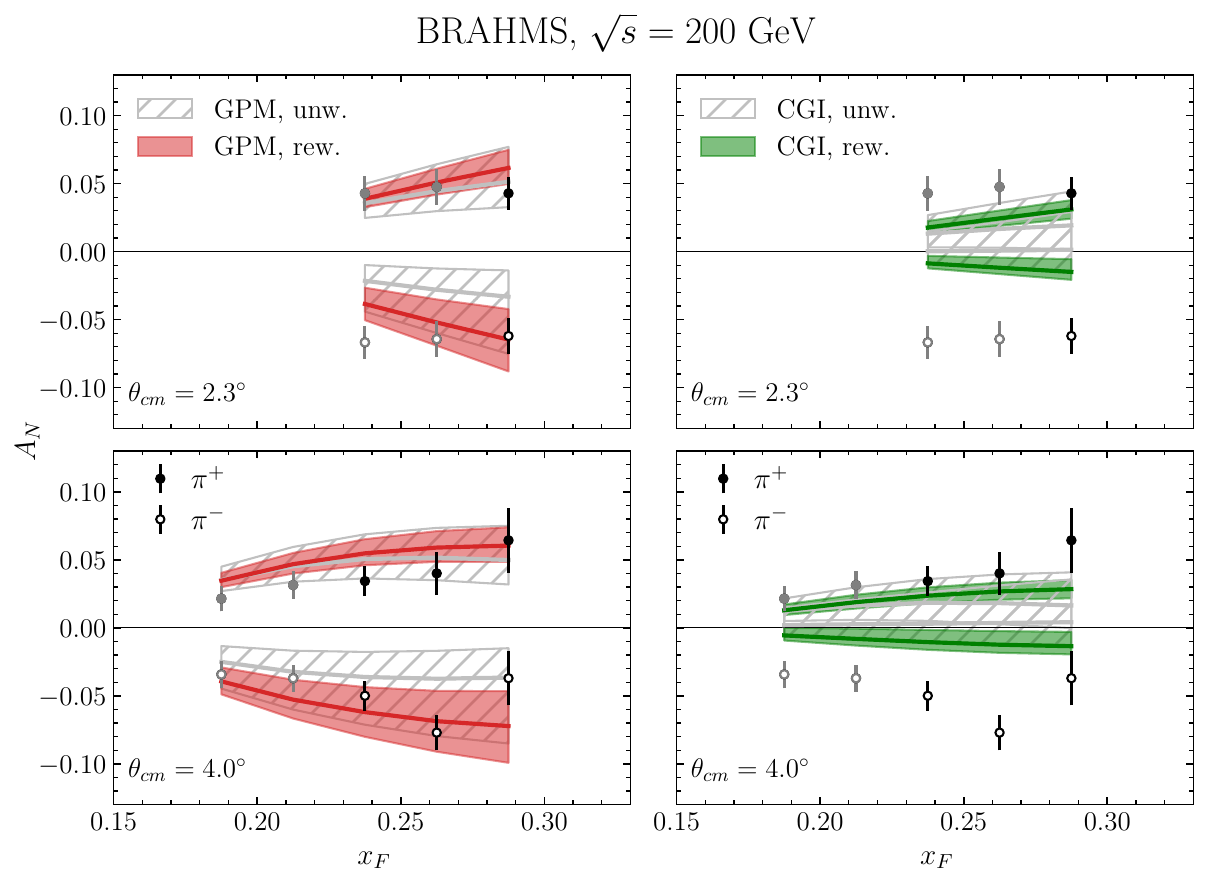}
\caption{Results for the simultaneous reweighting of the Sivers, transversity and Collins functions: unweighted and reweighted predictions for BRAHMS $A_N^{\pi^\pm}$ data~\cite{Lee:2007zzh} in the GPM (left panels) and the CGI-GPM (right panels) are presented. Data points in gray correspond to $P_T < 1.5$ GeV.}
\label{fig:AN_BRAHMS_rew_full}
\end{figure}

In Fig.~\ref{fig:AN_BRAHMS_rew_full} we show the unweighted and reweighted bands in the GPM and the CGI-GPM, compared to   $A_N$ data from BRAHMS for $\pi^+$ (full bullet points) and $\pi^-$ (empty bullet points). As expected, the reweighted curves present reduced uncertainties. The GPM describes these data better than the CGI-GPM, and the quality of the description increases if one does not consider the aforementioned data points with $P_T < 1.5$ GeV. A somehow larger discrepancy between our computation and the data is seen for $\pi^-$ in the CGI-GPM. 

\begin{figure}[t]
\centering
\includegraphics[width=\columnwidth]{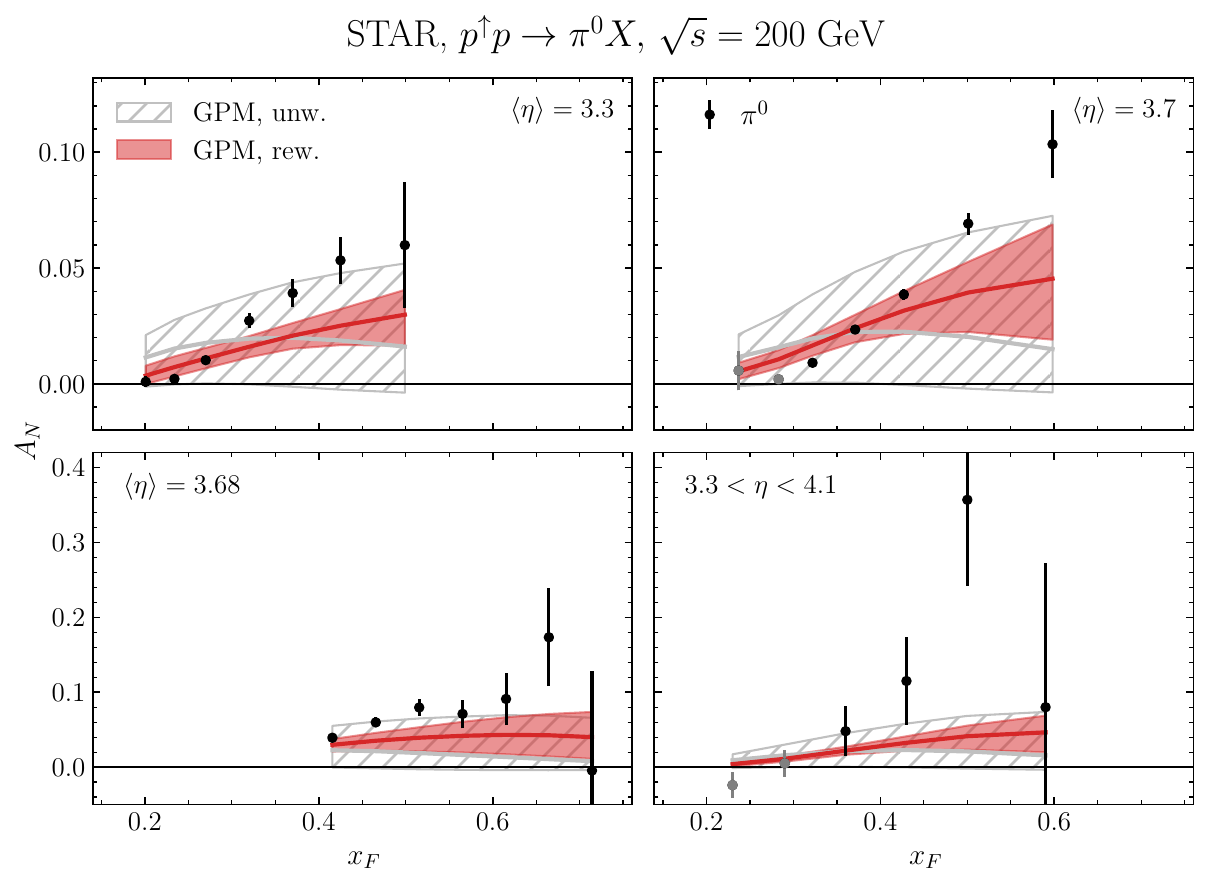}
\includegraphics[width=\columnwidth]{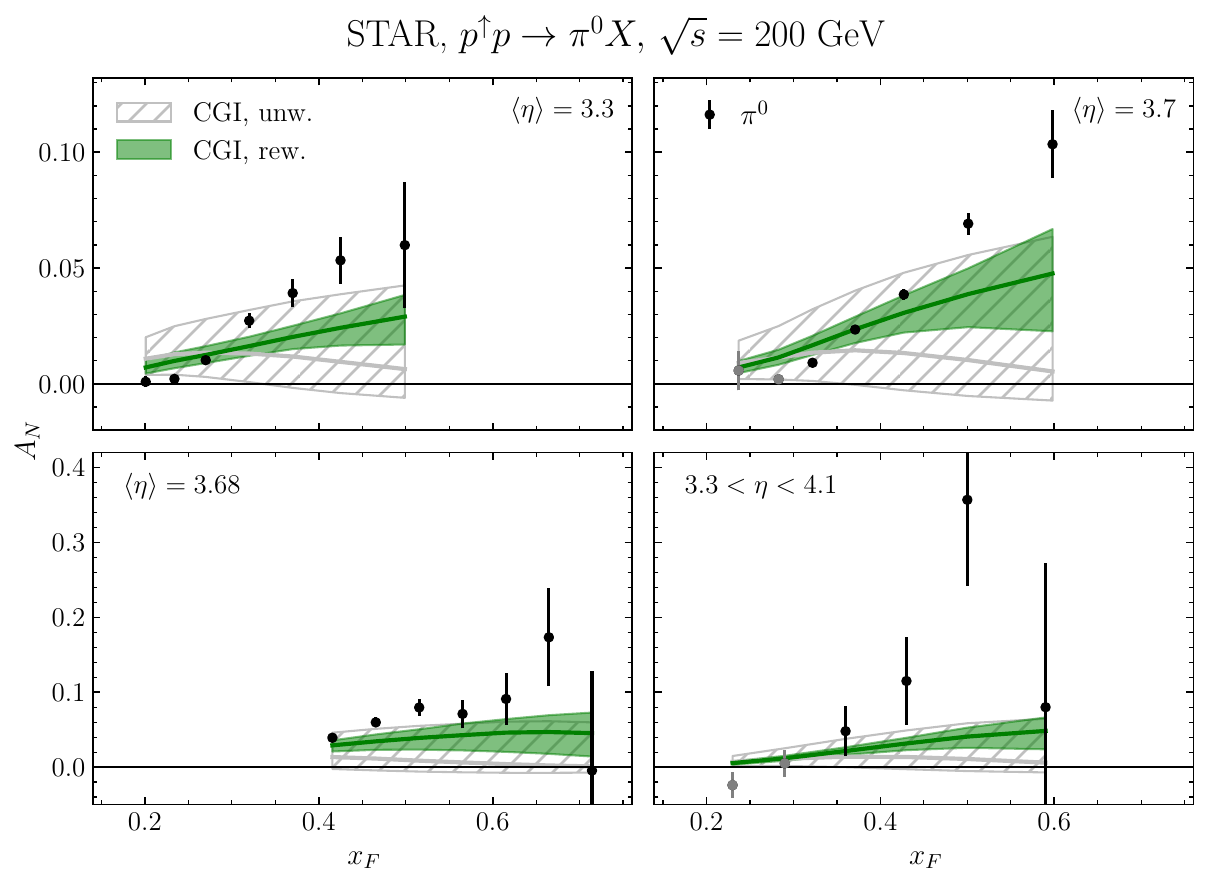}
\caption{Results for the simultaneous reweighting of the Sivers, transversity and Collins functions: reweighted curves for STAR data~\cite{STAR:2003lxu,STAR:2008ixi,STAR:2012ljf} in the GPM (upper panels) and the CGI-GPM (lower panels) are presented. Data points in gray correspond to $P_T < 1.5$ GeV.}
\label{fig:AN_STAR_old_rew_full}
\end{figure}

The comparison with the older STAR data~\cite{STAR:2003lxu,STAR:2008ixi,STAR:2012ljf}, collected without separating isolated and non-isolated pion samples, is shown in Fig.~\ref{fig:AN_STAR_old_rew_full} for the GPM (upper panels) and the CGI-GPM (lower panels), in four different ranges of pseudorapidity. Notice that, in the two kinematical configurations with largest $\langle \eta \rangle$ (right plots in the two panels), the first two data points at lower $x_F$ values correspond to $P_T < 1.5$ GeV. Both GPM and CGI-GPM estimates are in qualitative agreement with the data. The reweighted bands are able to describe the data at moderate $x_F$, and more interestingly, they present a shape that better represents the steady increase of the asymmetry at large-$x_F$ values, where the agreement is enhanced with respect to older analyses~\cite{Anselmino:2012rq, Anselmino:2013rya}. 

We finally move to the latest STAR data~\cite{STAR:2020nnl} for non-isolated $\pi^0s$. The kinematics of this dataset aligns more closely to that of our initial fits in SIDIS and $e^+e^-$, as it mainly involves pions with moderate momentum fractions $z$, excluding those with $z \sim 1$~\cite{STAR:2020nnl}. Furthermore, the $A_N$ data for non-isolated $\pi^0$ differ from the corresponding overall $\pi^0$ inclusive data sample, and from older $A_N$ measurements in similar kinematical regions~\cite{STAR:2003lxu,STAR:2008ixi,STAR:2012ljf}, as they do not show the usual pronounced steady increase at large $x_F$ (see also Figs.~6, 7 and 8 of Ref.~\cite{STAR:2020nnl} for a more exhaustive comparison). We will present the outcomes of the reweighting procedure, specifically addressing this STAR $\pi^0$ dataset, at the end of Section~\ref{sec:rew-impact} (omitting figures for brevity).

\begin{figure}[t]
\centering
\includegraphics[width=\columnwidth]{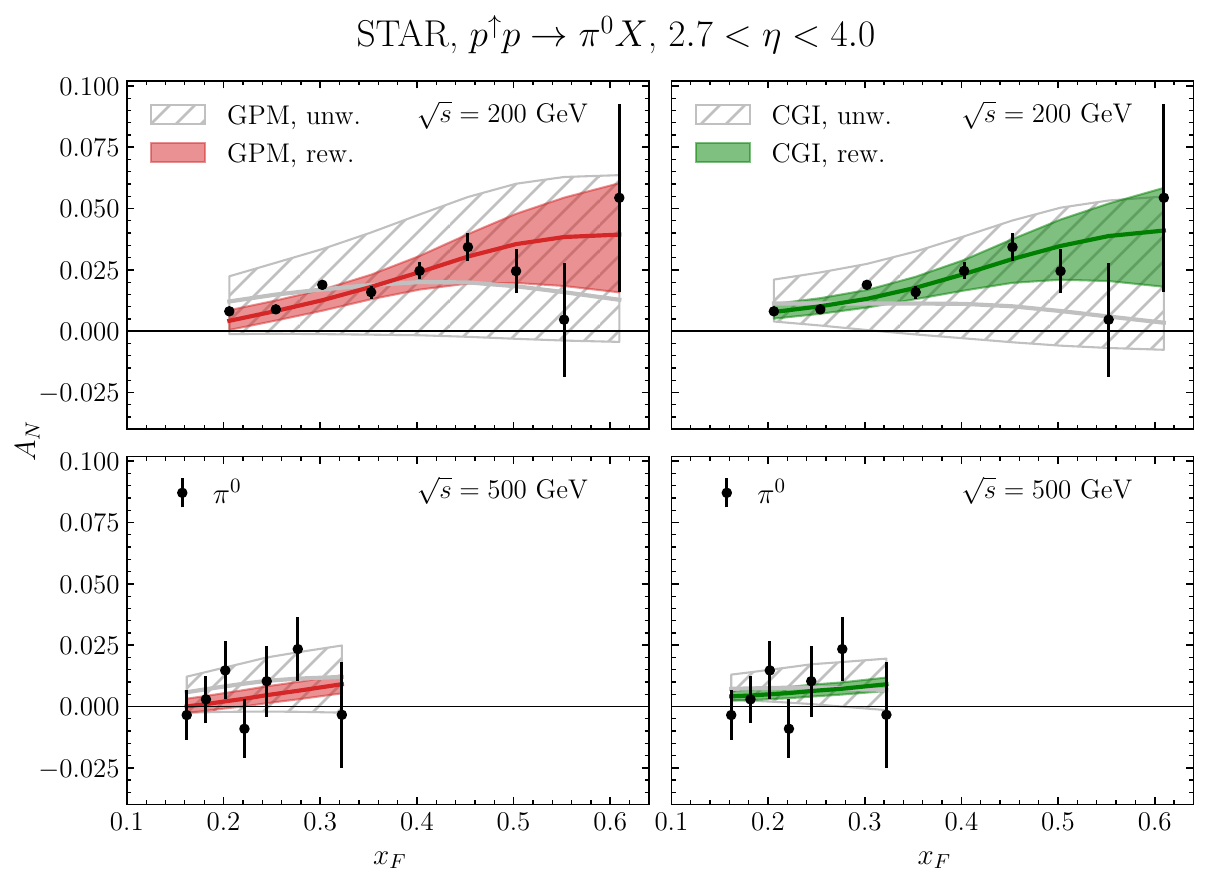}
\caption{Results for the simultaneous reweighting of the Sivers, transversity and Collins functions: unweighted and reweighted predictions of STAR $A_N$ data for non isolated neutral pions~\cite{STAR:2020nnl}. Comparisons of the asymmetries computed in the GPM (left panels) and in the CGI-GPM (right panels) with experimental data at $\sqrt{s} = 200$ GeV (upper panels) and $\sqrt{s} = 500$ GeV (lower panels) are presented. Here all data points correspond to $P_T > 1.5$ GeV.}
\label{fig:AN_STAR_rew_full}
\end{figure}

In Fig.~\ref{fig:AN_STAR_rew_full} we show our estimates and compare them against STAR results for non-isolated pions. Both GPM and CGI-GPM describe the data rather  well within uncertainties at the two different energies of $200$ and $500$ GeV. As the reweighting includes information from all the aforementioned datasets, we observe a steady increase at large $x_F$. However, when the reweighting is limited to the new STAR data alone, the shape of the reweighted bands appears flatter, mirroring the trend of the non-isolated pion data. In Section~\ref{sec:rew-impact}, we will also discuss the uncertainties affecting the TMDs and the corresponding $N_{\rm eff}$ obtained from reweighting in this specific case.

\subsection{Impact of $A_N$ data on Sivers, transversity and Collins functions \label{sec:rew-impact}}

We now examine the role played by $A_N$ data in the extraction of the Sivers, transversity and Collins functions. As a general feature, we anticipate that these data impact mostly on the TMD-PDFs, namely the Sivers and the transversity functions.

\begin{figure}[t]
\centering
\includegraphics[width=\columnwidth]{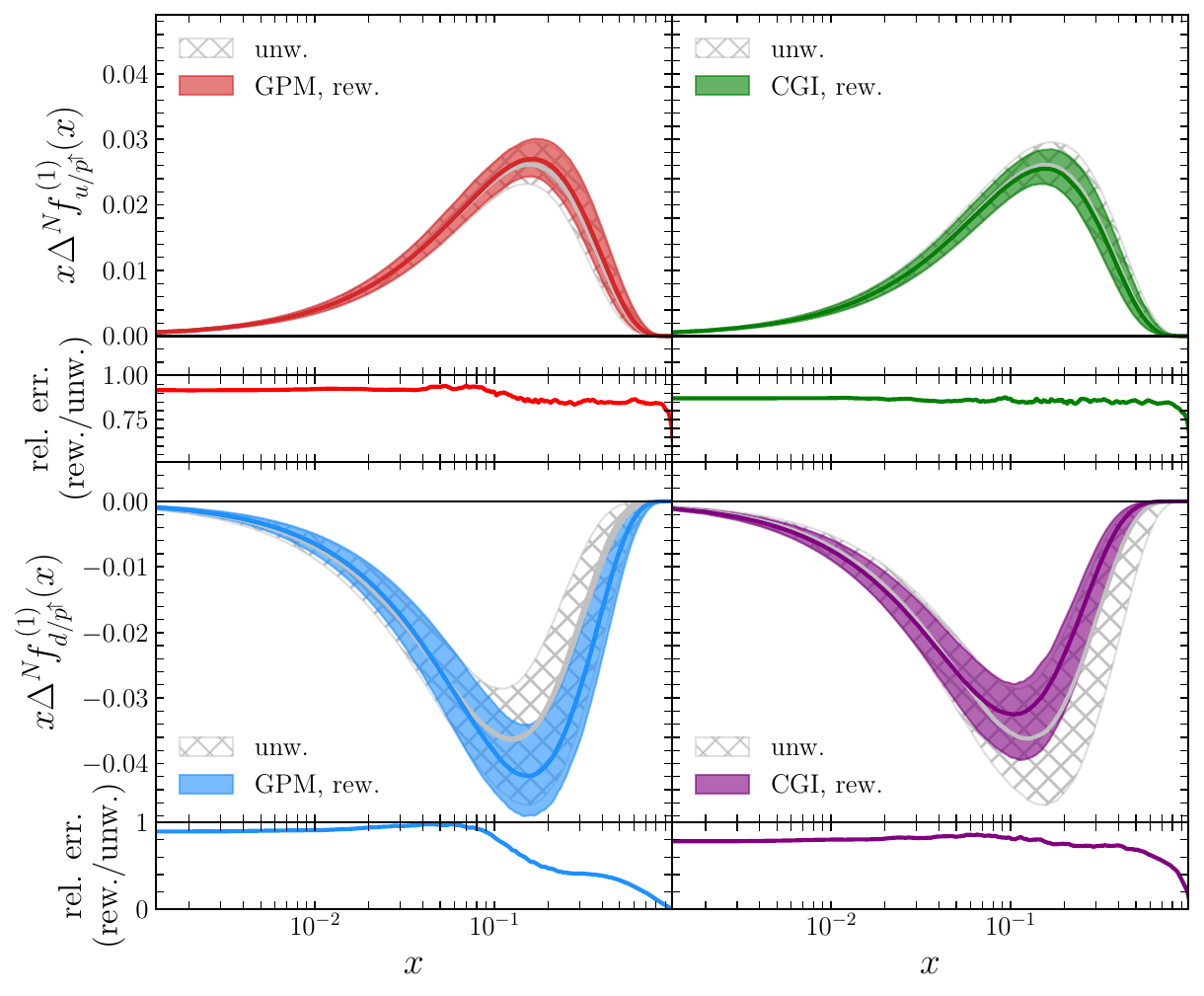}
\caption{Comparison of unweighted and reweighted first moments of up- (upper panels) and down-quark (lower panels) Sivers functions in the GPM (left panels) and in the CGI-GPM (right panels). The relative reduction of uncertainty is shown in the bottom panels.}
\label{fig:Sivers}
\end{figure}

\begin{figure}[h!]
\centering
\includegraphics[width=\columnwidth]{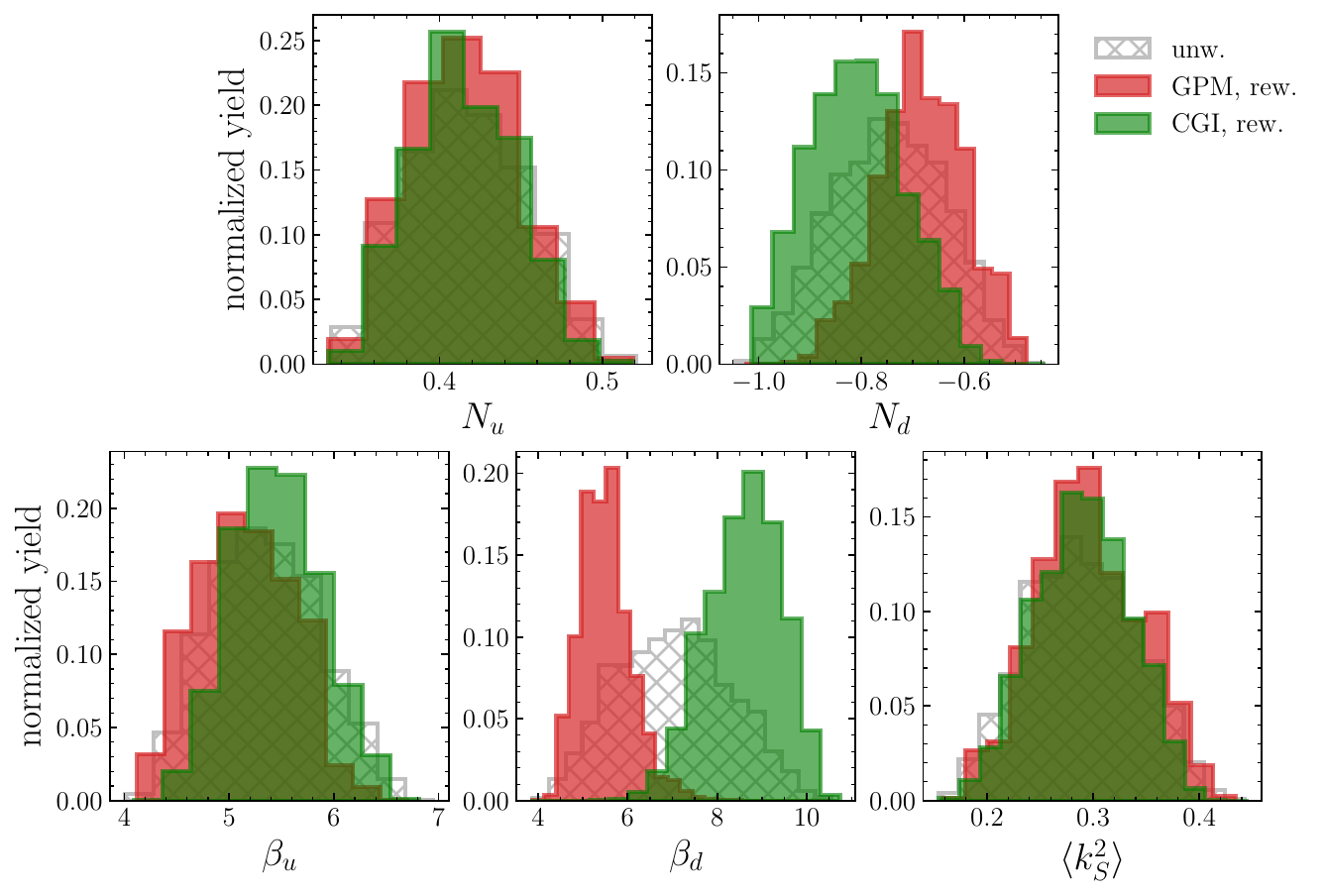}
\caption{Comparison of unweighted (gray) and reweighted distributions of parameters for the quark Sivers functions in the GPM (red) and in the CGI-GPM (green).}
\label{fig:Sivers-pars}
\end{figure}

We start by examining the Sivers case. In Fig.~\ref{fig:Sivers} we compare the unweighted and reweighted first moment of the quark Sivers functions, Eq.(\ref{eq:f1Tp-first-mom}), in the GPM (left panels) and CGI-GPM (right panels). As a general trend, the reweighted curves present reduced uncertainties. This reduction is more pronounced for the $d$-quark than for the $u$-quark Sivers function. The relative reduction of uncertainty of the reweighted Sivers first moments is about $20 - 30$\% for $f_{1T}^{\perp u}$ and $40 - 90$\% for $f_{1T}^{\perp d}$. The effective number of sets (see Eq.~(\ref{eq:N-eff})) surviving after reweighting is $N_{\rm eff} = 547\,(706)$ in the GPM (CGI-GPM) case. Fig.~\ref{fig:Sivers-pars} shows that, in both approaches, the parameters for the $u$-quark Sivers function and the Gaussian Sivers width do not change much, while the GPM appears to favor a smaller overall absolute value of the normalization for the $d$-quark Sivers function, with a slower decrease at large $x$ (smaller $\beta_d$ parameter), while the CGI seem to prefer a larger $N_d$ (in size), but with a faster decrease at large $x$.

Considering the transversity and Collins case, we emphasize that, though the Collins contribution to $A_N$ is formally the same in the GPM and CGI-GPM, the results for the reweighted curves for $h_1^q$ and $H_1^{\perp q}$ are slightly different. This reflects the different role of the Sivers contribution to $A_N$ in the two approaches.

\begin{figure}[h]
\centering
\includegraphics[width=\columnwidth]{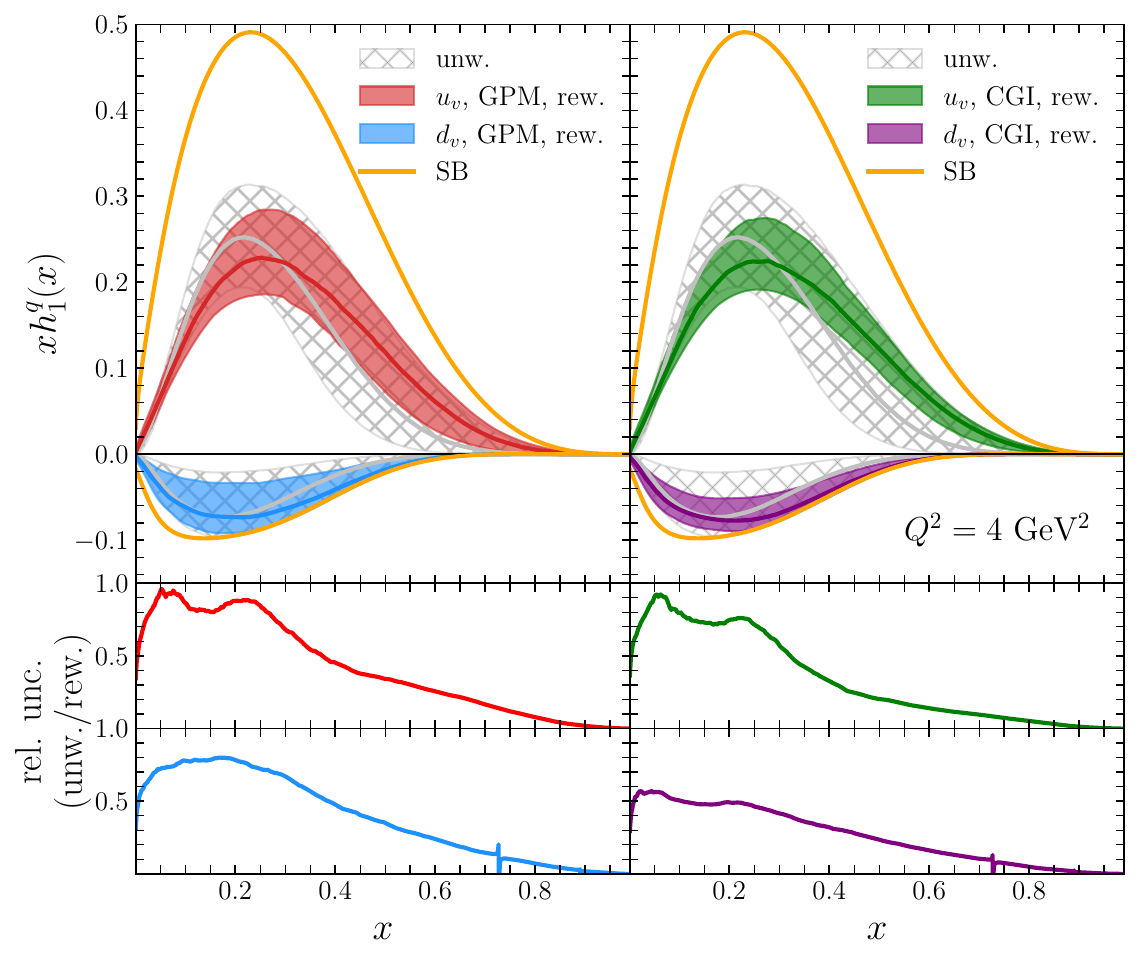}
\caption{Comparison of unweighted and reweighted $u_v$ and $d_v$ transversity functions in the GPM (left panels) and in the CGI-GPM (right panels). The corresponding Soffer bounds and the relative
reduction of uncertainty (same color coding in the bottom panels) are also shown.}
\label{fig:transversity}
\end{figure}

In Fig.~\ref{fig:transversity} we present the comparison between unweighted and reweighted $u_v$ and $d_v$ transversity functions, along with their corresponding Soffer bound, in the GPM (left panels) and CGI-GPM (right panels) at $Q^2 = 4$ GeV$^2$. Note that, compared to the unweighted results, $A_N$ data favor on average a slightly smaller $h_1^{u_v}$ in the region $x\lesssim 0.3$ and a slightly larger $h_1^{u_v}$ in the large-$x$ region. The inclusion of $A_N$ data sizeably reduces the uncertainty band in the region of $x\gtrsim 0.3$. As for $h_1^{d_v}$, we observe that a larger absolute value is preferred by the data on $A_N$. This is induced by the $A_N^{\pi^0}$ data at large $x_F$ (which are related to large $x$ values of the functions probed upon integration), that tend to favor sets yielding large asymmetries. The uncertainty reduction is about $20 - 30\%$ at smaller values of $x$, extending up to $80 -90\%$ at larger $x$ values for $h_1^{u_v}$, both in the GPM and in the CGI-GPM, while for $h_1^{d_v}$ the reduction is $30 -40\%$ (60\%) in the GPM (CGI-GPM) at small $x$ and up to $80 -90\%$ at large $x$ in both cases. Here, the effective number of sets after the reweighting is $N_{\rm eff} = 285$ (GPM) and $N_{\rm eff} = 110$ (CGI-GPM). This might be due to the poor description of $\pi^-$ data (see Fig.~\ref{fig:AN_BRAHMS_rew_full}). 

\begin{figure}[htbp]
\centering
\includegraphics[width=\columnwidth]{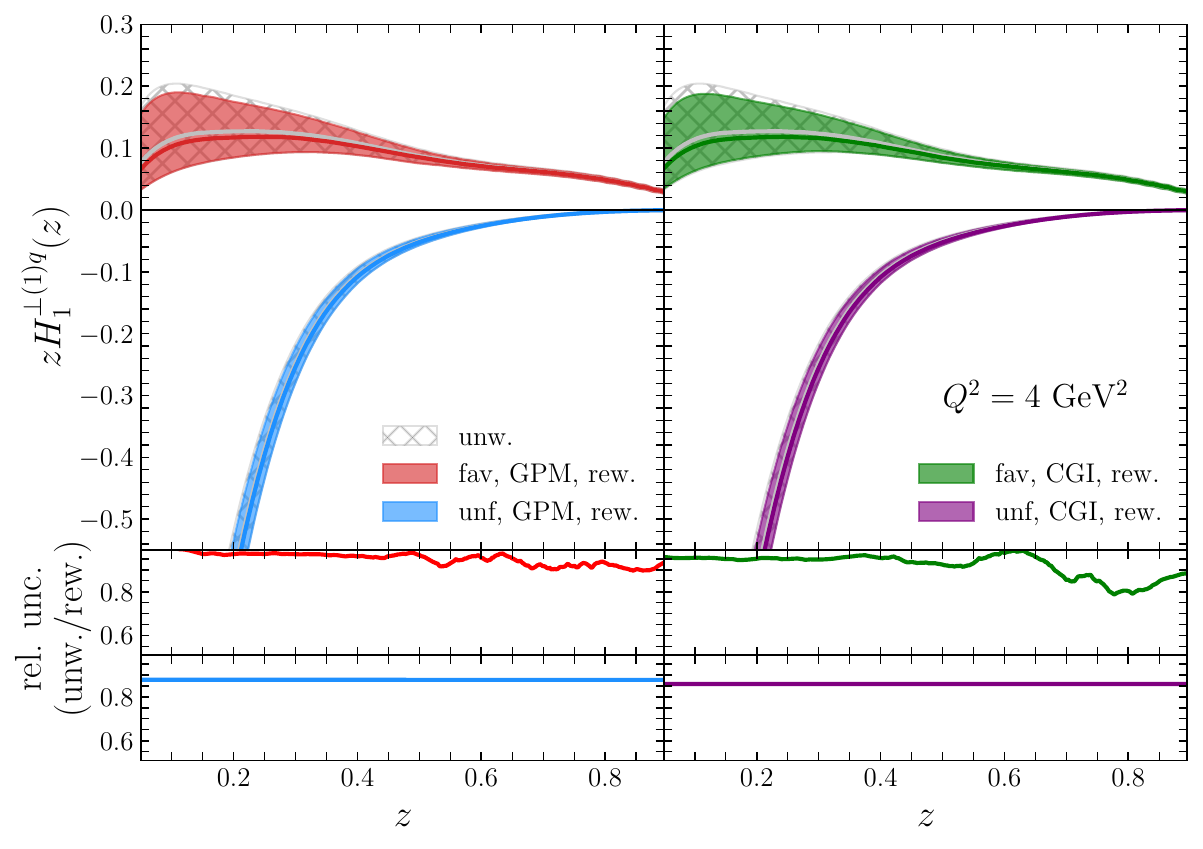}
\caption{Comparison of unweighted and reweighted favored (upper panels) and unfavored (lower panels) first moments of the Collins functions in the GPM (left panels) and in the CGI-GPM (right panels) at $Q^2 = 4$ GeV$^2$. The relative reduction of uncertainties is shown in the bottom plots.}
\label{fig:Collins}
\end{figure}

\begin{figure}[htbp]
\centering
\includegraphics[width=\columnwidth]{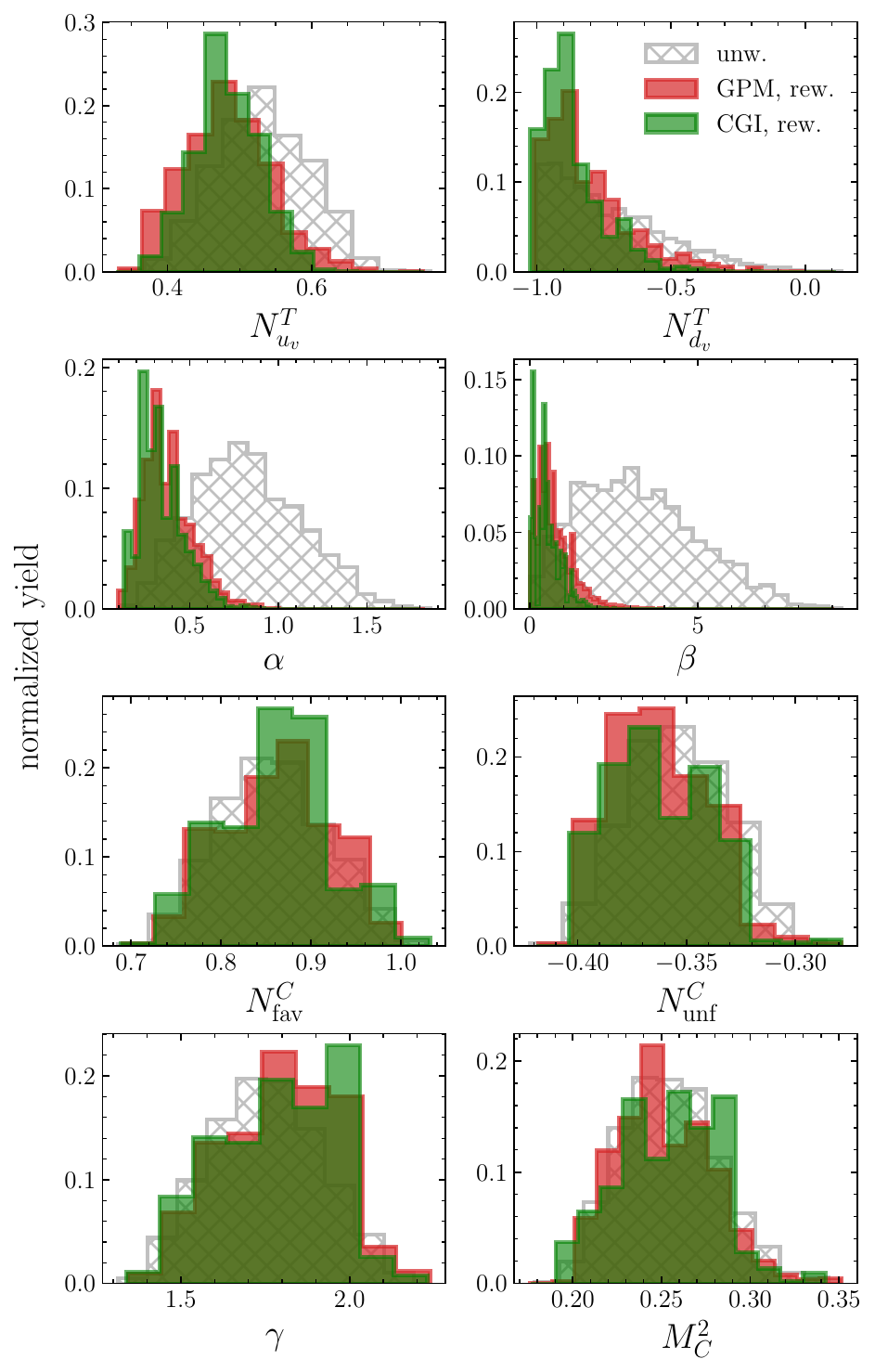}
\caption{Comparison of unweighted (gray) and reweighted distributions of parameters for the transversity and Collins extraction in the GPM (red) and in the CGI-GPM (green).}
\label{fig:Collins-pars}
\end{figure}

In Fig.~\ref{fig:Collins} we show the unweighted and reweighted Collins first moments in the two approaches, at $Q^2 = 4$ GeV$^2$, a typical SIDIS scale. This quantity is defined as~\cite{Meissner:2010cc} 
\begin{equation}\label{eq:H1perp-moment}
\begin{aligned}
    H_1^{\perp (1)\,q} (z) & = z^2 \int d^2 \bm{p}_\perp \frac{p_\perp^2}{2 m_h^2}\,H_1^{\perp q} (z, z^2 p_\perp^2) \\[1mm]
    & = \sqrt{\frac{e}{2}} \frac{1}{z m_h} \frac{M_C^3 \langle p^2_\perp \rangle}{\left(\langle p^2_\perp \rangle + M_C^2\right)^2}\,{\cal N}^C_q(z) D_{h/q}(z)\,,
    \end{aligned}
\end{equation}
where the last line is obtained adopting the pa\-ram\-e\-tri\-za\-tion in Eq.~\eqref{eq:Collins}. For these functions, the impact of $A_N$ data is less strong, but it allows for a reduction of the uncertainties (in both approaches) of about 5-10\% for the favored Collins function and about 15\% for the unfavored.

The previously mentioned slower decrease of the transversity function for increasing values of $x$ becomes evident when examining Fig.~\ref{fig:Collins-pars}, which compares unweighted (hatched gray histograms) and reweighted distributions of the fit parameters in both the GPM (red histograms) and CGI-GPM (green histograms). As noted above, $A_N$ data mainly affect the transversity function. This is clearly represented in the four top panels of Fig.~\ref{fig:Collins-pars}: reweighted values of $N^T_{u_v}$ tend to be smaller while the negative $N^T_{d_v}$ values are larger in size, approaching the limiting value of the Soffer bound ($\lvert N_q^T \rvert \leq 1$). At large $x$, $h_1^q$ tends to decrease following the Soffer bound rather closely (see the corresponding histogram of the $\beta$ parameter distribution, where the reweighted average values move close to zero). On the other hand, although the Collins parameter distributions show less sizeable variations, the uncertainties of the reweighted Collins functions are still slightly reduced. These considerations point towards the observation that the dominant contribution to $A_N$ is given by the Collins effect. This is consistent with some recent results obtained within the twist-3 approach~\cite{Cammarota:2020qcw, Gamberg:2022kdb}, where it was found that the main contribution to $A_N$ comes from the fragmentation mechanism.

Let us now briefly revisit the induced correlations that emerge from the simultaneous reweighting. We verified that the correlation matrix for the unweighted parameters factorizes into two submatrices (one for the $f_{1T}^\perp$ and one for the $h_1^q$ and $H_1^{\perp q}$ parameters). Conversely, as expected, some correlations are introduced by the reweighting procedure, as mentioned in our discussion on simultaneous reweighting. Specifically, we observe weak correlations between the Sivers and transversity normalizations and the $\beta$ parameters, both within the GPM and the CGI-GPM.

We finally provide a few remarks on the results we obtained for the reweighting procedure using solely the new STAR data for non-isolated $\pi^0$s. Interestingly, we note a more modest reduction in uncertainties of the reweighted TMDs, particularly for the Sivers and Collins functions, while for $h_1^q$ at higher $x$ values the reduction is more sizeable. Furthermore, we observe a higher effective number of retained sets, specifically $N_{\rm eff} = 1807\,(1961)$ for the Sivers fit within GPM (CGI-GPM), and $N_{\rm eff} = 1877$ and $1514$ for the transversity and Collins extractions within the GPM and CGI-GPM, respectively. These results appear to suggest a better compatibility between these latest STAR data and measurements from SIDIS and $e^+e^-$ experiments.

\begin{figure*}[ht!]
\centering
\includegraphics[width=.8\textwidth]{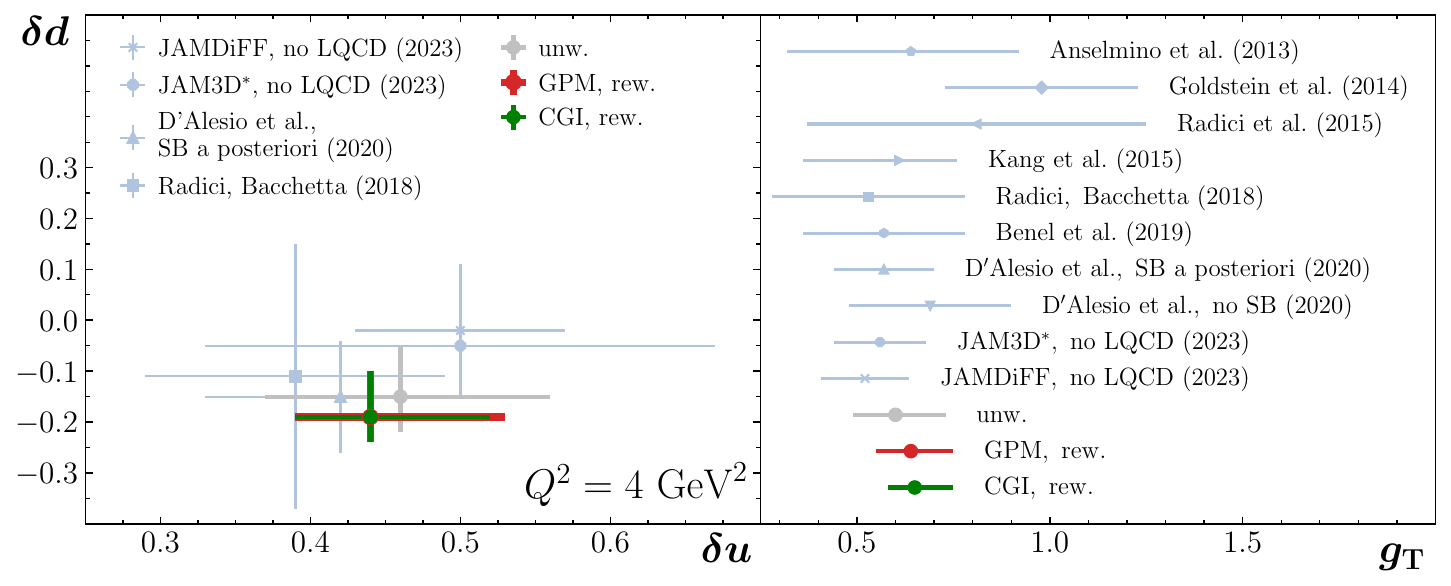}
\caption{Comparison of our results for the $u$ and $d$ tensor charges (left panel) and the iso-vector combination $g_T$ (right panel) with phenomenological estimates from Refs.~\cite{Anselmino:2013vqa, Goldstein:2014aja, Radici:2015mwa, Kang:2015msa, Radici:2018iag,  Benel:2019mcq, DAlesio:2020vtw, Cocuzza:2023oam} at $Q^2 = 4$ GeV$^2$. CGI-GPM (green) and GPM (red) reweighted central values almost coincide, having also similar uncertainties.}
\label{fig:gT-comparison}
\end{figure*}

\subsection{Tensor charges}

We conclude our analysis by reporting the corresponding values obtained for the nucleon tensor charges, defined as:
\begin{equation}
    \delta q 
    = \int\limits_0^1 \left[h_1^q(x) - h_1^{\bar q}(x)\right] dx\,, 
    \quad g_T  
    = \delta u - \delta d.
\end{equation}
The values obtained for the {\it unweighted} transversity functions at $Q^2 = 4$ GeV$^2$ are (central values are the median values): $\delta u = 0.46_{-0.09}^{+0.10}$, $\delta d = - 0.15_{-0.07}^{+0.10}$, $g_T = 0.60_{-0.11}^{+0.13}$. The {\it reweighted} tensor charges in the GPM (CGI-GPM) are: $\delta u = 0.47_{-0.07}^{+0.09}$ $(0.47_{-0.05}^{+0.08})$, $\delta d = - 0.18_{-0.06}^{+0.10}$ $(-0.19_{-0.05}^{+0.07})$, $g_T = 0.64_{-0.09}^{+0.11}$ $( 0.65_{-0.07}^{+0.10})$. These values are slightly larger as compared to older analyses (see Table 1 of Ref.~\cite{DAlesio:2020vtw}, ``using SB" case), due to the new HERMES data on proton, which render larger asymmetries and hence larger fitted functions, as previously observed, for instance, in Ref.~\cite{Gamberg:2022kdb}.

A detailed comparison of our results and various estimates of the tensor charges from phenomenological analyses is presented in Fig.~\ref{fig:gT-comparison}. We note that our current analysis and the majority of the previous studies of Refs.~\cite{Anselmino:2013vqa,Goldstein:2014aja,Radici:2015mwa,Kang:2015msa,Radici:2018iag,Benel:2019mcq,DAlesio:2020vtw, Cocuzza:2023oam} yield consistent values for $g_T$, $\delta u$, and $\delta d$. This corroborates the consistency of different extractions of transversity within different approaches exploiting a variety of experimental data.

\section{\label{sec:conclusions} Conclusions and outlook}

In this paper we have investigated the Bayesian reweighting procedure, extending it to the case of multiple, independent fits. For the first time, we have employed this technique to simultaneously reweight two independent extractions of quark TMD parton densities. Specifically, we have focused on the Sivers function and the TMD transversity and Collins functions. To this aim we have considered transverse single spin asymmetries data for inclusive pion production in polarized $pp$ collisions at RHIC.

The simultaneous reweighting involves two statistically independent fits, each with a substantially large MC sample ($\mathit{O}(10^5)$ sets) reflecting their corresponding uncertainty. Since these fits contribute additively to $A_N$, computing all possible combinations of both MC sets would in principle be necessary. However, to expedite the numerical computation and make them more efficient, we have developed and extended a compression technique for MC set samples. This innovative approach enabled us to exploit only 1\% of the sets in the reweighting process, without sacrificing any statistical information on parameter distributions. Such optimization not only enhances computational efficiency but also offers enough flexibility for further application to studies involving large sample parameter distributions.

Our phenomenological study, due to its peculiarities, has required an educated selection of the experimental data to be used for the reweighting procedure. The latest $A_N^{\pi^0}$ data from STAR Collaboration differ from previous measurements at RHIC, as they are provided separating non-isolated from isolated pions. The non-isolated dataset turned out to be more compatible with SIDIS and $e^+e^-$ measurements, for which TMD factorization holds and from which we extract the TMDs, \textit{i.e.}~our priors. Note however that, in our comprehensive analysis, we have included {\it all} available $A_N$ data for charged and neutral pions, obtaining a satisfactory global description.

The adopted dataset is dominated by $\pi^0$ production data, while only a few data points from BRAHMS for charged pions are available. As $A_N^{\pi^\pm}$ data are more sensitive to flavor separation, they could help in disentangling the issue of the predicted Sivers sign change. The description of these data seems to favor the GPM, where all TMDs are assumed to be universal and in which, contrary to the CGI-GPM, the expected Sivers sign change is not naturally recovered. The inclusion of data from future experiments, like COMPASS/AMBER~\cite{COMPASS:2010shj,Bradamante:2018ick,Adams:2018pwt}, JLab~\cite{Dudek:2012vr}, the EIC~\cite{Boer:2011fh,Accardi:2012qut} and the fixed-target programs at Tevatron at SpinQuest~\cite{SeaQuest:2019hsx}, and the LHC~\cite{Aidala:2019pit} would indeed be crucial in shedding light on this fundamental issue.

Our estimates exhibit improved agreement with data compared to previous analyses, owing partly to the careful incorporation of the Soffer bound in the TMD transversity distribution fit. This theoretical constraint, applied \textit{a posteriori}, renders larger asymmetries at large $x_F$, thereby favoring the dominance of the Collins mechanism, as observed in recent analyses within the collinear twist-3 formalism~\cite{Cammarota:2020qcw, Gamberg:2022kdb}.

Consistently with our previous work~\cite{Boglione:2021aha}, the reweighted TMD distributions present reduced uncertainties at large $x$, confirming once again the complementarity of $A_N$ data with SIDIS measurements. The reduction in uncertainty is about 40\% (90\%) for the $u$-($d$-)quark Sivers function, and about 80\% to 90\% for the $u_v$ and $d_v$ TMD transversity functions. The uncertainty reduction for the Collins functions is smaller (about 10\% for the favored and 20\% for the unfavored Collins TMDs), confirming that $e^+e^-$ data provide the strongest constraints on this polarized TMD fragmentation function.  

By performing the reweighting solely on non-isolated pion data, the reduction in uncertainties is smaller for the Sivers and Collins functions, and similar for the TMD transversity at large $x$. The retained $N_{\rm eff}$ sets in the GPM (CGI-GPM) case is $\sim 90\%$ ($\sim 95\%$) for the Sivers fit and $\sim 90\%$ ($\sim 75\%$) for the transversity and Collins extraction. This confirms an enhanced compatibility of SIDIS and $e^+e^-$ data with the new $A_N$ measurements from STAR for non-isolated pions.

This work is a natural extension of our previous study~\cite{Boglione:2021aha}, and a proof of concept for upcoming TMD analyses. Future studies will explore different TMD parametrizations and incorporate new data from COMPASS/AMBER~\cite{Bradamante:2018ick}, JLab~\cite{Dudek:2012vr}, and from the future Electron-Ion Collider~\cite{Boer:2011fh,Accardi:2012qut}. Additionally, planned investigations into inclusive jet or pion-in-jet production data in $pp$ collisions,  where Sivers and Collins effects can be accessed individually, will further contribute to a more comprehensive understanding of TMD dynamics, universality and factorization breaking effects.

\section*{Acknowledgments}

We thank Mauro Anselmino for numerous discussions. We are grateful to Emanuele Nocera and Andrea Signori for discussions on the statistical approaches and technical aspects of the current analysis. C.F.~and A.P.~are grateful to the Physics Department of the University of Cagliari and the INFN Cagliari division for the kind hospitality extended to them during the development of this project. A.P.~acknowledges financial support from the University of Cagliari under the Visiting Professor Programme. U.D.~also acknowledges financial support from Fondazione di Sardegna under the project ``Matter-antimatter asymmetry and polarisation in strange hadrons at LHCb'', project number F73C22001150007 (University of Cagliari). This article is part of a project that has received funding from the European Union's Horizon 2020 research and innovation programme under grant agreement STRONG-2020 - No.~824093. This work was partially supported by the Grant for Internationalization SIGA\_GFI\_22\_01 financed by Torino University, by the National Science Foundation under Grants  No.~PHY-2012002, No.~PHY-2310031, No.~PHY-2335114 (A.P.), and by the U.S. Department of Energy contract No.~DE-AC05-06OR23177, under which Jefferson Science Associates, LLC operates Jefferson Lab (A.P.).

\end{document}